\documentclass[%
 reprint,
nofootinbib,
 amsmath,amssymb,
 aps,
]{revtex4-2}

\usepackage{graphicx}
\usepackage{dcolumn}
\usepackage{bm}

\usepackage{url}
\usepackage{hyperref}

\usepackage[percent]{overpic}
\usepackage{tikz}
\usepackage{caption}
\usepackage{graphicx}
\usepackage{color}
\usepackage[normalem]{ulem}
\usepackage{subcaption}
\usepackage{graphicx}
\usepackage{comment}

\newcommand{\ac}{\hat{\mathcal{C}}}
\begin{document}

\preprint{APS/123-QED}

\title{Ultracompact Anisotropic Stars and Gravastars in General Relativity}

\author{Prajwal Hassan Puttasiddappa}
 \email{prajwal.puttasiddappa@edu.ufes.br}
\affiliation{PPGCosmo, Universidade Federal do Esp\'irito Santo, 29075-910, Vit\'oria, ES, Brazil}%
\affiliation{Departamento de F\'isica Te\'orica, Instituto de F\'isica, Universidade do Estado de Rio de Janeiro, CEP 20550-013, Rio de Janeiro, Brazil}
\affiliation{Institute of Theoretical Astrophysics, University of Oslo, Sem Sælands vei 13, 0371 Oslo, Norway}
\author{Nora Bretón}
\affiliation{Departmento de Física, Centro de Investigación y de Estudios Avanzados del Instituto Politécnico Nacional (Cinvestav), PO. Box 14-740, Mexico City, Mexico}

\author{Santiago Esteban Perez Bergliaffa}
\affiliation{Departamento de F\'isica Te\'orica, Instituto de F\'isica,
Universidade do Estado de Rio de Janeiro,
CEP 20550-013, Rio de Janeiro, Brazil
}

\date{\today}

\begin{abstract}
We study static, spherically symmetric ultracompact objects in General Relativity with anisotropic stress. Using the covariant anisotropic equation of state and assuming homogeneous density, we construct a mostly analytical model admitting configurations beyond the Buchdahl limit and with compactness arbitrarily close to the black hole value. The solutions exhibit two regimes separated by a critical curve in the parameter space of compactness and anisotropy, associated with the divergence of the central pressure: regular anisotropic configurations with positive central pressure, and singular configurations with negative central pressure for arbitrary anisotropy. In the latter case, we show that introducing a thick shell removes the pressure divergence, yielding regular ultracompact gravastar configurations. We further construct both anisotropic and isotropic gravastar models using this thick-shell construction.
\end{abstract}

\maketitle


\section{\label{sec:Intro}Introduction:}

Compact objects are equilibrium solutions of the Einstein equations describing the final stages of gravitational collapse. Among the known astrophysical compact objects, neutron stars represent the most compact equilibrium configurations sustained by ordinary matter, reaching compactness $M/R \lesssim 0.3$. General Relativity also admits black holes, which correspond to the limiting case of gravitational collapse and are described by stationary vacuum solutions with compactness $M/R  = 1/2$. Black holes are distinguished by two defining geometric features: an event horizon and a spacetime singularity. While singularity theorems predict the formation of spacetime singularities through geodesic incompleteness under rather general conditions, their presence signals a breakdown of the classical theory. At the semiclassical level, the presence of an event horizon gives rise to unresolved issues such as the black hole information paradox. These considerations naturally motivate the question of whether General Relativity admits equilibrium configurations that can approach the compactness of a black hole while remaining free of an event horizon and spacetime singularities \cite{Iyercompact1985, elastic}.

The increasing precision of gravitational-wave observations and horizon-scale imaging has transformed this question from a purely theoretical one into an observationally testable problem \cite{Maggio2021, Vagnozzi2022, Cardoso2014, Cardoso2019}. These developments have stimulated interest in ultracompact, horizonless objects that may arise within General Relativity, provided suitable matter configurations. Such objects can approach the compactness of a black hole and remain externally indistinguishable from black holes, while possessing nontrivial matter configurations in their interiors.

The possibility of constructing such objects within GR is constrained by fundamental bounds on compactness. In particular, for isotropic perfect fluids, Buchdahl's theorem \cite{Buchdahl1959} establishes the maximum compactness $M/R = 4/9$, which is below the Schwarzschild black hole value. To obtain configurations with higher compactness, the assumptions entering this bound must be relaxed. For example, more general matter sources within General Relativity can be considered, such as boson stars, where additional matter degrees of freedom generate compact equilibrium configurations \cite{Liebling2012}, or the underlying theory of gravity can be modified \cite{Bueno2025}. 

In this work, we consider anisotropic stress within GR, which provides a particularly simple and physically motivated way to surpass the Buchdahl limit \cite{Andreasson2007}. Since the pioneering work \cite{Bowers1974}, it has been realized that anisotropy can arise from several sources (for a review see \cite{Herrera1997}), such as superfluidity, strong magnetic fields, and viscosity (for a list of references, see \cite{Lau2024}). Beyond these sources of anisotropy in the static configuration, dissipative processes like heat flow, viscous dissipation and shear, which are generically present during gravitational collapse and stellar evolution, drive an initially isotropic fluid toward an anisotropic pressure distribution, and there is no a priori reason for this acquired anisotropy to vanish once equilibrium is reached \cite{Herrera2020}. Recent gravitational-wave observations have also enabled constraints on anisotropic effects in compact objects \cite{Guedes2025}. 

The properties of such configurations depend on the choice of equation of state (EOS) relating the radial and tangential pressures. Here, we consider a covariant EOS containing a parameter that directly controls the anisotropic pressure, which was introduced in \cite{Raposo2018}, together with the assumption of homogeneous density. This allows us to obtain a mostly analytical model of static, spherically symmetric configurations. The resulting solutions include both regular and singular configurations, separated by a critical curve determined by the anisotropy parameter and compactness. In particular, we show that beyond-Buchdahl compactness can be achieved in two distinct regimes: large anisotropy with positive central pressure, and arbitrary anisotropy with negative central pressure, the latter corresponding to a gravastar configuration \cite{Cattoen2005,Jampolski2023,Jampolski2025}.

The structure of the paper is the following. In Section \ref{sec:aniso_stars}, the equations that describe the anisotropic model are presented, the regular and singular configurations are characterized, and the main features of the former are discussed. Regular anisotropic configurations with negative central pressure (namely, gravastars) are examined in Section \ref{gravastars}, as well as their isotropic limit.  We present in Section \ref{concl} our closing remarks.

\section{\label{sec:aniso_stars}Anisotropic uniform stars}

Consider a generic static, spherically symmetric distribution of fluid with energy density $\rho(r)$, radial pressure $P_r(r)$, and tangential pressure $P_t(r) \neq P_r(r)$, described by the stress-energy tensor

\begin{equation}\label{stressTmunu}
    T_{\mu\nu} = (\rho + P_t) u_\mu u_\nu + P_t g_{\mu\nu} - \Delta k_\mu k_\nu \ ,
\end{equation}

where $u^\mu$ is the fluid 4-velocity ($u^\mu u_\mu = -1$) and $k^\mu$ is a unit spacelike vector ($k^\mu k_\mu = +1$) orthogonal to $u^\mu$ ($u^\mu k_\mu = 0$). The pressure anisotropy is defined as $\Delta(r) \equiv P_t(r) - P_r(r)$. For the spherically symmetric line element,

\begin{equation}\label{ansatzmetric}
    ds^2 = -e^{A(r)} dt^2 + e^{B(r)}dr^2 + r^2 (d\theta^2 + \sin^2\theta d\varphi^2)\ ,
\end{equation}

with $e^{-B(r)} = 1 - \frac{2m(r)}{r}$, the Einstein's field equations lead to
\footnote{We use units such that $c = 1 = G$.}
,
\begin{equation}
\label{toveqs}
\begin{split}
    \frac{e^{-B}}{r^2}\left(\frac{2m(r)}{r - 2 m(r)} + rB'\right) &= 8\pi \rho\ ,\\
    \frac{e^{-B}}{r^2}\left(rA'(r) - \frac{2m(r)}{r - 2m(r)}\right) &= 8\pi P_r\ ,\\
    \frac{e^{-B}}{4r}\left( rA'^2 - 2B' + A'(2 - rB') + 2rA''\right) &= 8\pi  P_t\ ,
\end{split}
\end{equation}
where a
prime indicates derivative w.r.t $r$. 

The first of these equations integrates to give the mass function,
\begin{equation}\label{sol1}
    m(r) = 4\pi \int_0^r \rho(\bar{r})\bar{r}^2\,d\bar{r}\ .
\end{equation}
At the stellar surface $r = R$, where $\rho = 0$, the mass function reduces to the mass of the star $M = m(R)$. Making use of the Bianchi identity, the conservation of the stress-energy tensor ($\nabla^\mu T_{\mu\nu} = 0$) gives the Tolman-Oppenheimer-Volkoff (TOV) equation\footnote{Note that $P_r$, but not $P_t$, enters the gravitational force term, reflecting the distinct roles of the two principal stresses in the relativistic hydrostatic equilibrium equation \cite{Herrera2020}.}:

\begin{equation}\label{genericTOV}
    P_r'(r) + \frac{A'(r)}{2}(\rho + P_r) - \frac{2}{r}\Delta(r) = 0\ .
\end{equation}

The metric function $A(r)$ is determined from the second field equation:
\begin{equation}\label{Aeq}
    A'(r) = \frac{2\left(m(r) + 4\pi r^3 P_r\right)}{r^2 \left(1 - \frac{2m(r)}{r}\right)}\ .
\end{equation}

The system comprising Eqs.\eqref{sol1}, \eqref{genericTOV} and \eqref{Aeq}  has five unknown functions of $r$, namely, $\{ \rho\ , P_r\ , P_t\ , A\ , B\}$. Hence, two equations of state may be specified to close it. We choose to specify $\rho(r)$ and $\Delta(r)$.

The equations must be supplemented by
adequate physical boundary conditions. We require that:
\begin{itemize}
    \item[-] At the center ($r \to 0$), $m(r) \to 0$, and the radial pressure displays an extremum, hence $P_r'(r) = 0$. Consequently, from Eq. \eqref{genericTOV}, $A'(r) \to 0$. 
    
    \item[-] 
    The stellar surface is defined by $P_r(R)=0$.
    In contrast, the tangential pressure $P_t(r)$ need not vanish at $r=R$.

    \item[-] Spherical symmetry demands that the fluid anisotropy must vanish at least as rapidly as $r$ when $r\to 0$ \cite{Bowers1974}. 

\end{itemize}

In this work, we study the consequences of
the covariant EOS proposed in \cite{Raposo2018}:
\begin{equation}\label{tangpressansatz}
     \Delta(r) =  \mathcal{C} f(\rho) k^\mu \nabla_\mu P_r\ ,
\end{equation}
where $\mathcal{C}\geq 0$ is a dimensionful parameter controlling the degree of isotropy with $\mathcal{C}=0$ corresponding to the isotropic limit, while $f(\rho)$ is a function of the matter density. This form of equation of state naturally couples the anisotropic pressure to the radial pressure gradient and guarantees that $\Delta(r)$ vanishes at the center of symmetry. For other (non-covariant) choices, see \cite{Bowers1974,Dev2000}. 

Substituting Eq.~\eqref{tangpressansatz} into the TOV equation leads to,

\begin{equation}\label{TOVfull}
    P_r'(r) = - \frac{ (m(r) + 4\pi r^3 G P_r)(P_r + \rho) }{ r(r - 2m(r)) \left(1 + \frac{2\mathcal{C} f(\rho)}{r} \sqrt{1 - \frac{2m(r)}{r}} \right)}\ ,
\end{equation}
which reduces to the standard isotropic TOV equation when $\mathcal{C}=0$. In what follows, we adopt the simple choice $f(\rho)=\rho$, and assume a constant density profile, $\rho=$ constant. The case $f(\rho)=\rho$ with a polytropic equation of state was investigated numerically in \cite{Raposo2018}.

\subsection{\label{sec:homo_stars}Homogeneous stars}

We now consider the special case of an incompressible fluid with constant density $ \rho_*$. In this case, the function $f(\rho_*) \equiv \rho_*$ can be absorbed into the constant $\mathcal{C}$, and therefore we may set $f(\rho_*) = 1$ without loss of generality. From the mass equation \eqref{sol1}, the mass function is trivially integrated and the total mass of the star becomes $m(R) \equiv M = \rho_* (4\pi R^3/3)$. Under these assumptions, Eq.\eqref{Aeq} becomes
\begin{equation}\label{eqAinH}
     A'(r) = \frac{r\left(\frac{2M}{R^3} + 8\pi P_r\right)}{1 - \frac{2Mr^2}{R^3}}\ .
\end{equation}
Substituting this expression into the modified TOV equation \eqref{TOVfull}  yields the differential equation for the radial pressure,
\begin{equation}
    P'_r(r) = \frac{r^2(\frac{2M}{R^3} + 8\pi GP_r)(P_r + \rho_*)}{2\left(1 - \frac{2M}{R^3}r^2\right)\left(r + 2\mathcal{C} \sqrt{1 - \frac{2M}{R^3} r^2}\right)}\ .
\end{equation}

To simplify the analysis, we render the variables and parameters dimensionless, and use the notation $\hat{r} = r/R$, $\hat{M} = M/R$, $\ac = \mathcal{C}/R$, etc. The stellar configurations are then characterized by the dimensionless parameters $(\hat{M}, \ac)$, and $\hat r$, with $\hat{\rho}_* =\frac{3\hat M}{4\pi}$, and the surface of the star located at $\hat r=1$.

The radial pressure admits an analytical solution of the form, 
\begin{equation}\label{radialpressure}
    \hat{P}_r (\hat{r})= \hat{\rho}_* \left(\frac{F(\hat{M},\hat{r}, \ac) - 1}{1 - 3F(\hat{M},\hat{r},\ac)}\right)\ ,
\end{equation}
where the function $F(\hat{M}, \hat{r}, \ac)$ is given by, 

\begin{widetext}
\begin{equation}\label{Fexpression}
     F = \sqrt{\frac{1 - 2\hat{M}}{1 - 2\hat{M}\hat{r}^2}}
    \left( \frac{ \hat{r} + 2 \ac\sqrt{1 - 2\hat{M} \hat{r}^2 }}{ 1 + 2\ac\sqrt{1 - 2\hat{M}}} \right)^{\frac{8\hat{M} \ac^2}{1 + 8\hat{M} \ac^2}} \\   
    \exp\left(\frac{2\ac \sqrt{2\hat{M}}}{1 + 8 \hat{M} \ac^2}\left(\arcsin\left(\sqrt{2\hat{M}}\right) - \arcsin \left(\sqrt{2\hat{M}} \hat{r}\right)\right)\right) \ ,
\end{equation}
\end{widetext}

The surface boundary condition $\hat{P}_r(\hat{r} =1)=0$ is automatically satisfied since $F(\hat{M},1,\ac)=1$. In the isotropic limit $\ac = 0$, this expression reduces to the pressure profile of the interior Schwarzschild star solution
\cite{Schwarzschild1916},

\begin{equation}\label{isotropicPr}
    \hat{P}^{\rm iso}_r = \hat{\rho}_*\left(\frac{\sqrt{1 - 2\hat{M}} - \sqrt{1 - 2\hat{M}\hat{r}^2}}{\sqrt{1 - 2\hat{M}\hat{r}^2} - 3 \sqrt{1 - 2\hat{M}}}\right)\ .
\end{equation}
Of particular interest is the central radial pressure, $\hat{P}_c \equiv \hat{P}_r(\hat{r} = 0)$:

\begin{widetext}
\begin{equation}\label{Pceq}
    \hat{P}_c(\hat{M}, \ac) = -\hat{\rho}_*\frac{\left(2\ac\sqrt{1-2\hat{M}} + 1\right)^{\frac{8\ac^2\hat{M}}{1 + 8\ac^2\hat{M}}} - 
    \sqrt{1-2\hat{M}} (2\ac)^{\frac{8\ac^2\hat{M}}{1 + 8\ac^2\hat{M}}}
    e^\frac{2\sqrt{2\hat{M}}\ac\arcsin(\sqrt{2\hat{M}})}{8\ac^2\hat{M} + 1}}
    {\left(2\ac\sqrt{1-2\hat{M}} + 1\right)^{\frac{8\ac^2\hat{M}}{1 + 8\ac^2\hat{M}}} - 
    3\sqrt{1-2\hat{M}} (2\ac)^{\frac{8\ac^2\hat{M}}{1 + 8\ac^2\hat{M}}}
    e^\frac{2\sqrt{2\hat{M}}\ac\arcsin(\sqrt{2\hat{M}})}{8\ac^2\hat{M} + 1}}\ ,
\end{equation}
\end{widetext}

In the isotropic limit $\ac = 0$, the divergence of the central pressure reproduces the compactness bound derived by Buchdahl \cite{Buchdahl1959}, namely $\hat{M} < \hat{M}_B = \frac{4}{9}$. This inequality guarantees the central pressure is finite and non-negative throughout the object.

\begin{figure}
    \centering
    \includegraphics[width=\linewidth]{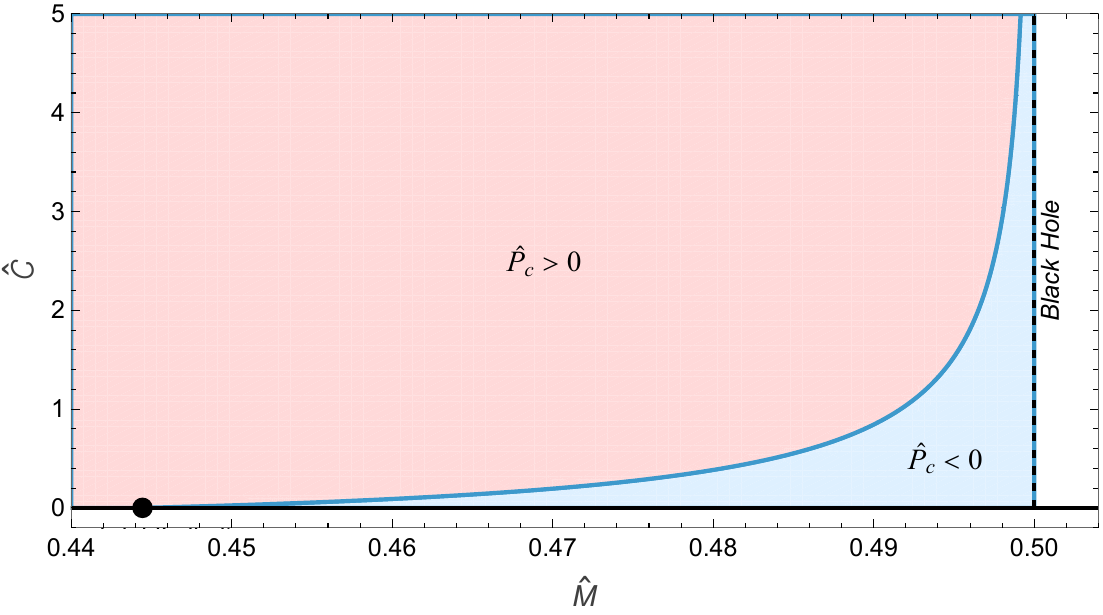}
    \caption{The critical (thick blue) line indicates the relation between the anisotropy parameter $\ac$ and the compacticity $\hat{M}$ that follows from imposing that the denominator of $\hat{P}_c$ in \eqref{Pceq} vanishes, leading to a divergence in the central pressure. The line separates the configurations with positive and negative central pressure in the $(\ac,\hat{M})$ plane. The red region shows the parameter space where one could have regular stars with compactness beyond the Buchdahl bound (indicated as $\bullet$) with an everywhere positive pressure inside the star. The blue region corresponds to configurations with negative central pressure. The black dot indicates Buchdahl's limit.}    \label{fig:CMdivergecontour}
\end{figure}

When anisotropy is included ($\ac>0$), the divergence of the radial pressure at the origin (which leads to a divergence of the tangential pressure via Eq.\eqref{tangpressansatz}) must be re-examined. The vanishing of the denominator of Eq.\eqref{Pceq} defines a critical curve in the $(\hat{M}, \ac)$ plane shown in Fig.\ref{fig:CMdivergecontour}. The curve separates the parameter space into configurations with positive and negative central pressure represented in red and blue, respectively. Hence, there are regular configurations with compactness beyond the Buchdahl bound, with an everywhere positive pressure inside the star even for mild anisotropy. For large values of $\ac$, the compactness approaches that of a black hole. The same figure also reveals a class of solutions with compactness arbitrarily close to the black hole value with a negative central pressure for any $\ac$. These configurations display a divergence of the radial pressure, which may be regularized to obtain a gravastar configuration, and will be discussed in Sec.\ref{gravastars}. 

Using the anisotropic EOS \eqref{tangpressansatz}, the tangential pressure if found to be,
\begin{equation}
\begin{split}
    \hat{P}_t &= \frac{ \hat{\rho}_*}{(1 - 3F)^2} \Bigg[F\Bigg(4 - 3F \\ &\quad + \frac{4\ac\hat{M}\hat{r}^2}{\sqrt{1 - 2\hat{M}\hat{r}^2}\left(\hat{r} + 2\ac\sqrt{1 - 2\hat{M}\hat{r}^2}\right)} \Bigg) - 1\Bigg]\ .
    \label{ptang}
\end{split}
\end{equation}

\begin{figure*}
    \centering
    
    \begin{subfigure}{0.329\textwidth}
        \centering
        \includegraphics[width=\linewidth]{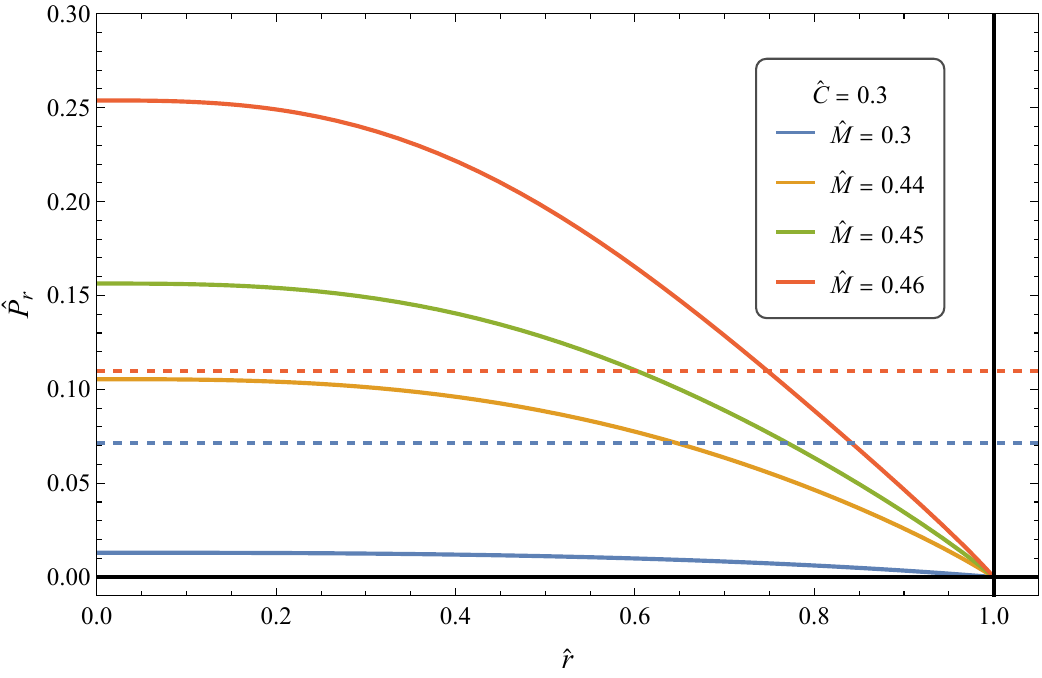}
        \caption{$\ac = 0.3$}
        \label{fig:radial_pressureregular}
    \end{subfigure}
    \hfill
    \begin{subfigure}{0.329\textwidth}
        \centering
        \includegraphics[width=\linewidth]{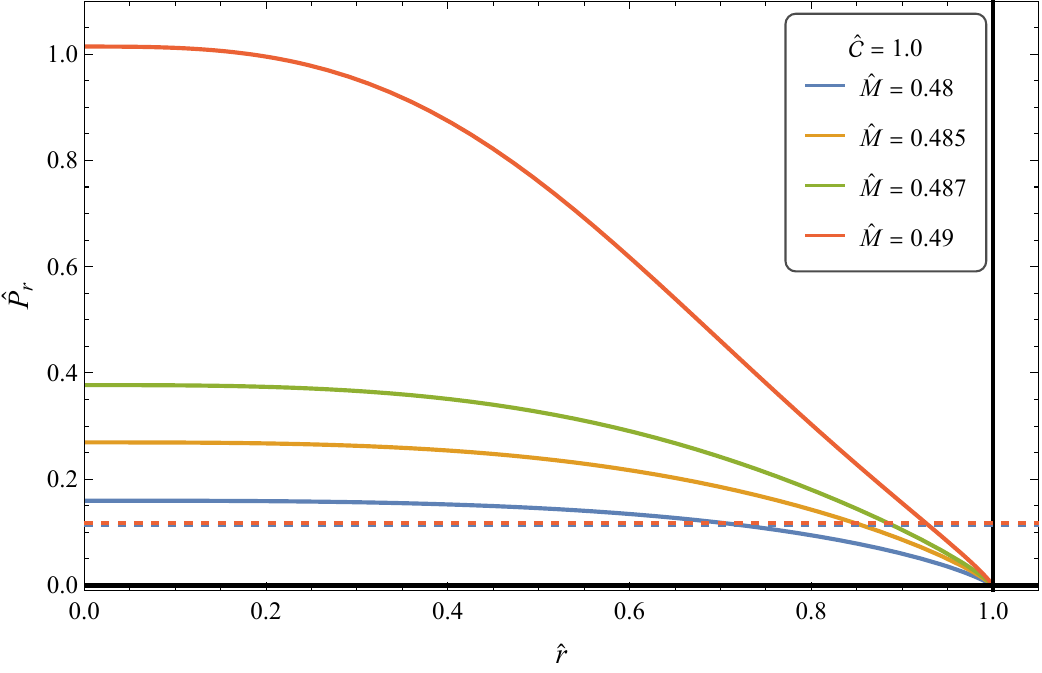}
        \caption{$\ac = 1.0$}
        \label{fig:radialpressureregular1}
    \end{subfigure}
    \hfill
    \begin{subfigure}{0.329\textwidth}
        \centering
        \includegraphics[width=\linewidth]{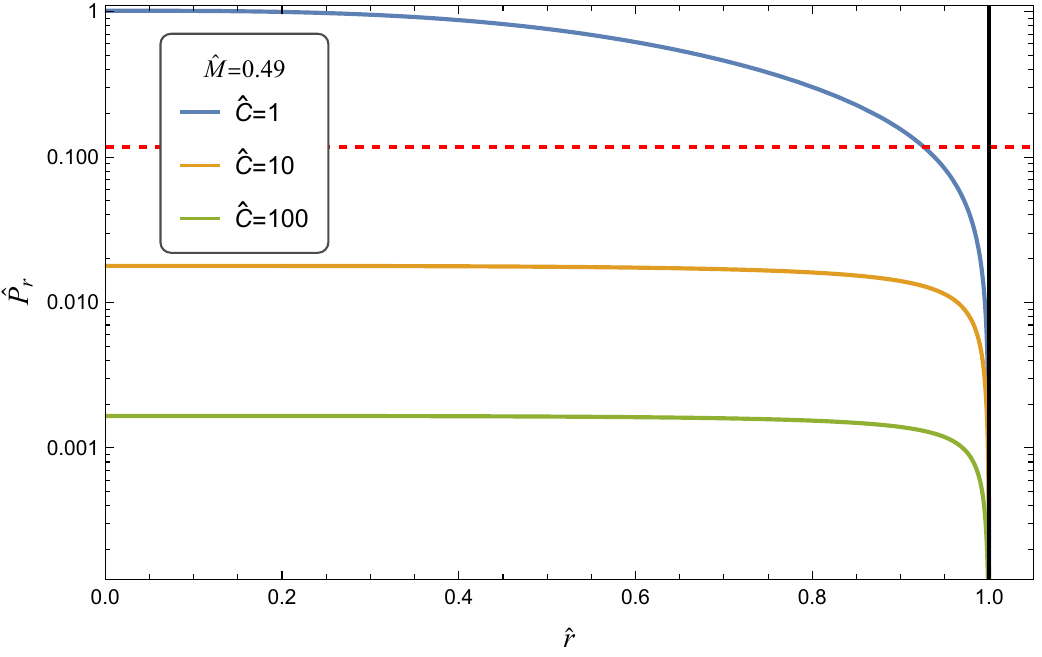}
        \caption{$\hat{M} = 0.49$}
        \label{fig:dec_pr}
    \end{subfigure}
    
    \caption{Figure (a) (respectively, (b)) shows radial pressure profiles for $\ac=0.3$ ($\ac=1.0$) for different values of $\hat M$. Figure (c) shows the radial pressure for $\hat M = 0.49$ and several values of $\ac$. The horizontal dashed lines show the value of $\rho_*$ for the minimum and maximum of the chosen values of $\hat M$.}
\label{fig:rad_profiles1}
\end{figure*}

\begin{figure*}
    \centering
    
    \begin{subfigure}{0.329\textwidth}
        \centering
        \includegraphics[width=\linewidth]{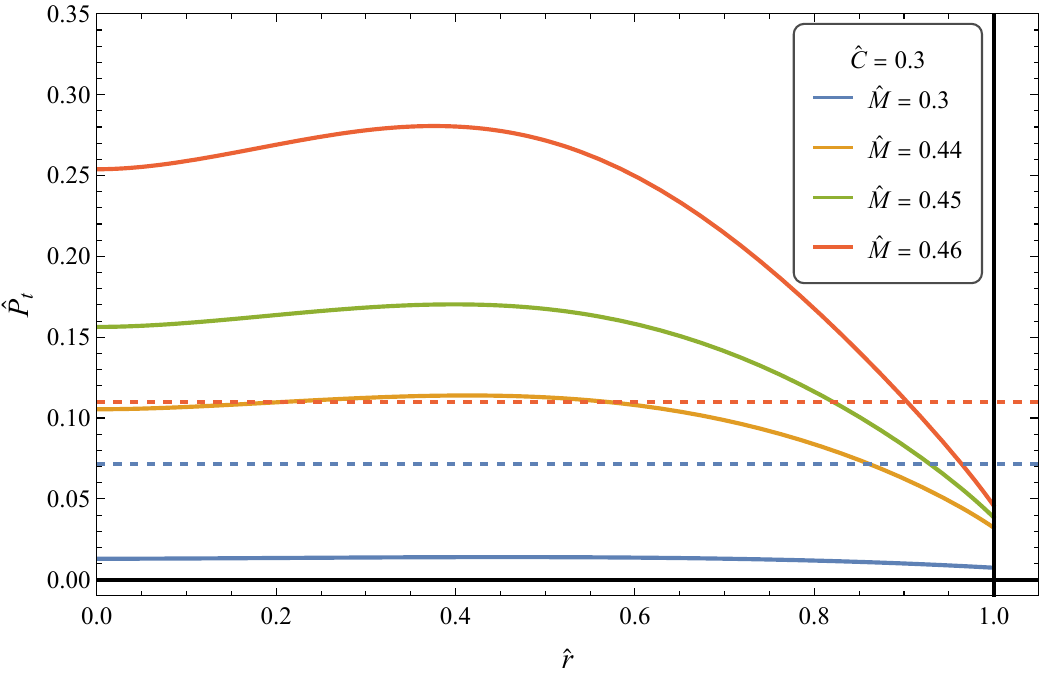}
        \caption{$\ac = 0.3$}
        \label{fig:tangentialpressregular}
    \end{subfigure}
    \hfill
    \begin{subfigure}{0.329\textwidth}
        \centering
        \includegraphics[width=\linewidth]{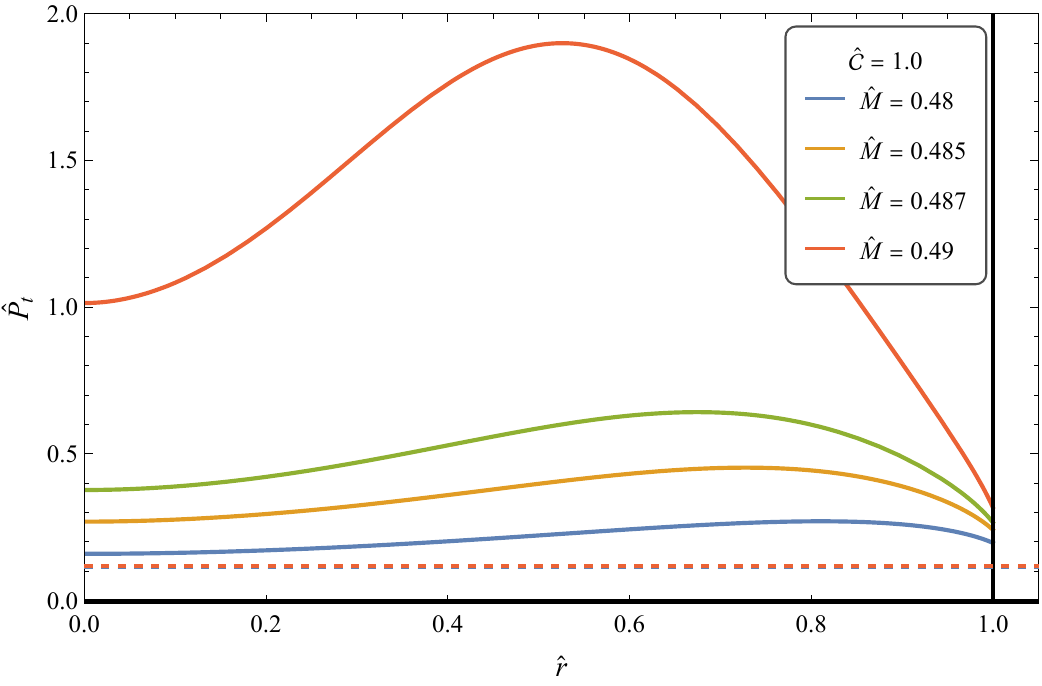}
        \caption{$\ac = 1.0$}
        \label{fig:tangentialpressregular1}
    \end{subfigure}
    \hfill
    \begin{subfigure}{0.329\textwidth}
        \centering
        \includegraphics[width=\linewidth]{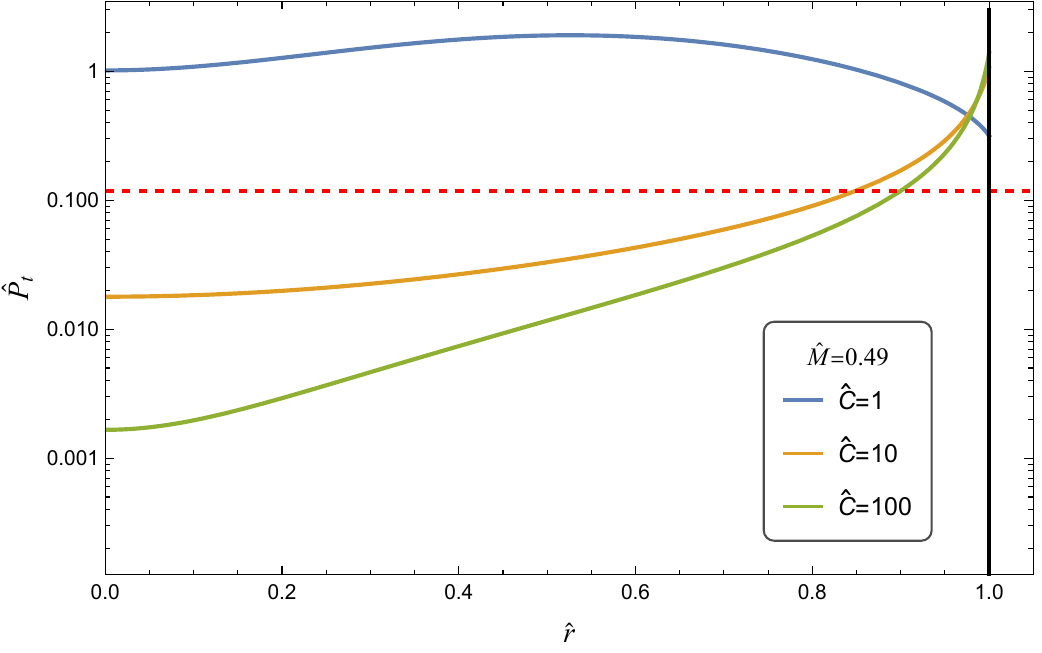}
        \caption{$\hat{M} = 0.49$}
        \label{fig:dec_pt}
    \end{subfigure}
    
    \caption{Tangential pressure profiles, 
    following the same details as those in Fig.\ref{fig:rad_profiles1}}
\label{fig:tang_profiles1}
\end{figure*}

Radial and tangential pressure profiles are shown in Figs.\ref{fig:rad_profiles1} and \ref{fig:tang_profiles1}, respectively, for representative values of the anisotropy parameter and different values of $\hat M$ corresponding to regular configurations. For any $\ac$, $\hat P_r$ is a monotonically decreasing function of $\hat r$, while $\hat P_t$ displays a maximum for low values of $\ac$. For high values of $\ac$, the tangential pressure increases monotonically from the center, and displays a maximum at the surface of the star. The plots show ultra-compact stellar configurations with finite and everywhere positive  $\hat{P}_r>0$ and moderate and high values of $\ac$, in accordance with Fig.\ref{fig:CMdivergecontour}. Taking the limit of very large $\ac$, it follows that $F\rightarrow 1$, $\hat P_r\rightarrow 0$ (see Fig.\ref{fig:dec_pr}) and $\hat P_t\propto Mr^2/(1-2Mr^2)$. In this case, for $\hat M$ very close to 0.5, the tangential pressure is negligible inside the star except near the surface, as seen in Fig.\ref{fig:dec_pt}.

The metric function $A(r)$ is obtained from Eq.\eqref{eqAinH}, which, using the results presented above, can be written as, 
\begin{multline}\label{gtt}
    A(\hat{r})  = 4 \hat{M} \int^{\hat{r}}_{1} \frac{\hat{r}' d\hat{r}'}{(3F(\hat{r}') - 1) (1 - 2\hat{M} \hat{r}'^2)}\\  + \ln (1 -2\hat{M})\ .
\end{multline}

The integration constant 
in $A$ is fixed by matching the interior solution to the exterior Schwarzschild metric at the stellar surface. The metric coefficient $g_{tt}=e^A$ is well-behaved and everywhere positive (as can be seen from Fig.\ref{fig:effectivepotential}).

\subsubsection{Energy conditions}

Before the examination of the standard pointwise energy conditions for the anisotropic regular configurations, we study the behaviour of the function $F(\hat{r},\hat{M},  \ac)$, which determines the pressures. From Eq.\eqref{radialpressure} it follows that the radial pressure $\hat{P}_r(\hat{r})$ is positive and finite for
\begin{equation}\label{SEC}
    1\geq F(\hat{M},\hat{r},\ac)>\frac{1}{3}\ .
\end{equation}

\begin{figure}
    \centering
    \includegraphics[width=0.99\linewidth]{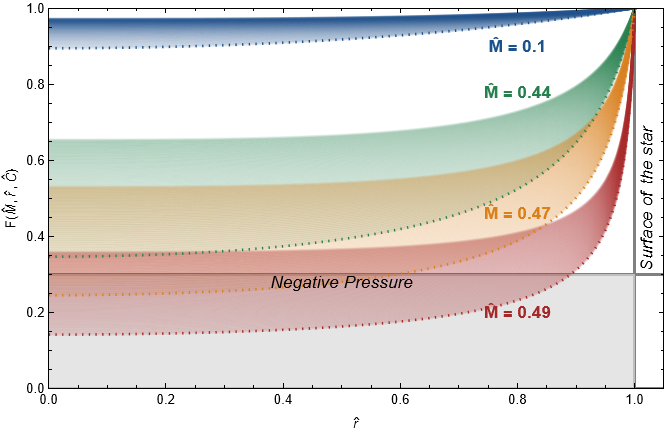}
    \caption{The figure shows the plot of $F(\hat{r})$ for $\hat{M} = 0.1, 0.44,0.47,0.49$ and $\ac \in [0, 1]$. For each value of $\hat M$, the different values of $\ac$ define a region, with the lower limit (corresponding to $\ac = 0$) shown in dotted curves. The curves in yellow and red show that, for a sufficiently high $\ac$, there are configurations for which the pressure is everywhere positive. For lower values of $\ac$ and high $\hat M$, the radial pressure can take negative values. Such configurations are singular, but they can be regularized to describe a gravastar (see Sect.\ref{gravastars}).}
    \label{fig:energycondition}
\end{figure}

At the surface of the star, $F$ reaches unity for any $\ac$, ensuring that $\hat{P}_r(1) = 0$. When $F < 1/3$, the radial pressure becomes negative, while the limiting value $F = 1/3$ corresponds to the divergence of the central pressure, see Fig.\ref{fig:energycondition}. The lower bound is essential for regular anisotropic stars without pressure divergences. As shown previously, these regular stars can be ultra-compact, \textit{i.e.} have compactness beyond the Buchdahl limit, reaching values close to that of a black hole. The tangential pressure, follows from Eq.\eqref{ptang}; for $\ac \geq 0$, and positive radial pressure, $\hat{P}_t \geq \hat{P}_r$. Consequently, both pressures remain positive throughout the stellar interior within the bound $F>1/3$. 

The energy conditions are most conveniently evaluated in the comoving orthonormal frame, where the stress-energy tensor is diagonal. For the anisotropic fluid described by the stress-energy tensor in Eq.\eqref{stressTmunu}, the null energy condition (NEC) leads to the two inequalities $\hat{\rho}_* + \hat{P}_r(\hat{r}) \geq 0$ and $\hat{\rho}_* + \hat{P}_t(\hat{r}) \geq 0$. Since the density is positive and both pressures remain non-negative within the star, these conditions are automatically satisfied. The strong energy condition (SEC) requires $\hat{\rho}_* + \hat{P}_r(\hat{r}) + 2\hat{P}_t (\hat{r}) \geq 0$. As argued above, $\hat{P}_t\geq \hat{P}_r\geq 0$. Hence, the SEC is satisfied for any value of $\hat r$.

Finally, the dominant energy condition (DEC) imposes stronger requirements: $\hat{\rho}_* \geq |\hat{P}_r(\hat{r})|$ and $\hat{\rho}_* \geq |\hat{P}_t(\hat{r})|$. As shown in Figs. \ref{fig:radial_pressureregular} and \ref{fig:tangentialpressregular}, DEC is satisfied for very low $\ac$ and $\hat M\lessapprox \hat M_B$. Figs.\ref{fig:radialpressureregular1} and \ref{fig:tangentialpressregular1} show that DEC is violated for $\mathcal{O}(1)$ anisotropy and high compactness. It is seen from Figs.\ref{fig:dec_pr} and \ref{fig:dec_pt} that DEC violation is restricted to occur very near the surface for configurations with high $\hat M$ and $\ac$, due to the rise of $\hat P_t$ there, in line with the findings in \cite{Raposo2018}.

\subsubsection{Null geodesics and trapping zones}

As discussed in \cite{Abramowicz1999,Cardoso2014,Stuchlik2017}, the existence of a region in which null geodesics are trapped can be relevant for gravitational waves and is related to the stability of the object. Since the trapping zones follow from the effective potential, we present its main features for the regular anisotropic configurations presented above. The corresponding analysis for the isotropic case was studied in \cite{Stuchlik2001}.

The motion of null particles in a general spherically symmetric spacetime described by the metric \eqref{ansatzmetric}, restricted to the equatorial plane ($\theta = \pi/2$), is governed by,
\begin{equation}
    \left(\frac{e^{A+B}}{L^2}\right) \dot{r}^2 = \frac{1}{b^2} - V(r)  \ ,
\end{equation}
where the effective potential is $V(r) \equiv \frac{e^{A(r)}}{r^2}$ and the impact parameter is given by $b = \frac{L}{E}$. Turning points of the trajectory are determined by $b^{-2} = V(r)$. Circular null geodesics correspond to extrema of the effective potential: maxima correspond to unstable null orbits, while minima correspond to stable circular null geodesics. 

\begin{figure}
    \centering
\begin{subfigure}{0.5\textwidth}
    \centering
    \includegraphics[width=\linewidth]{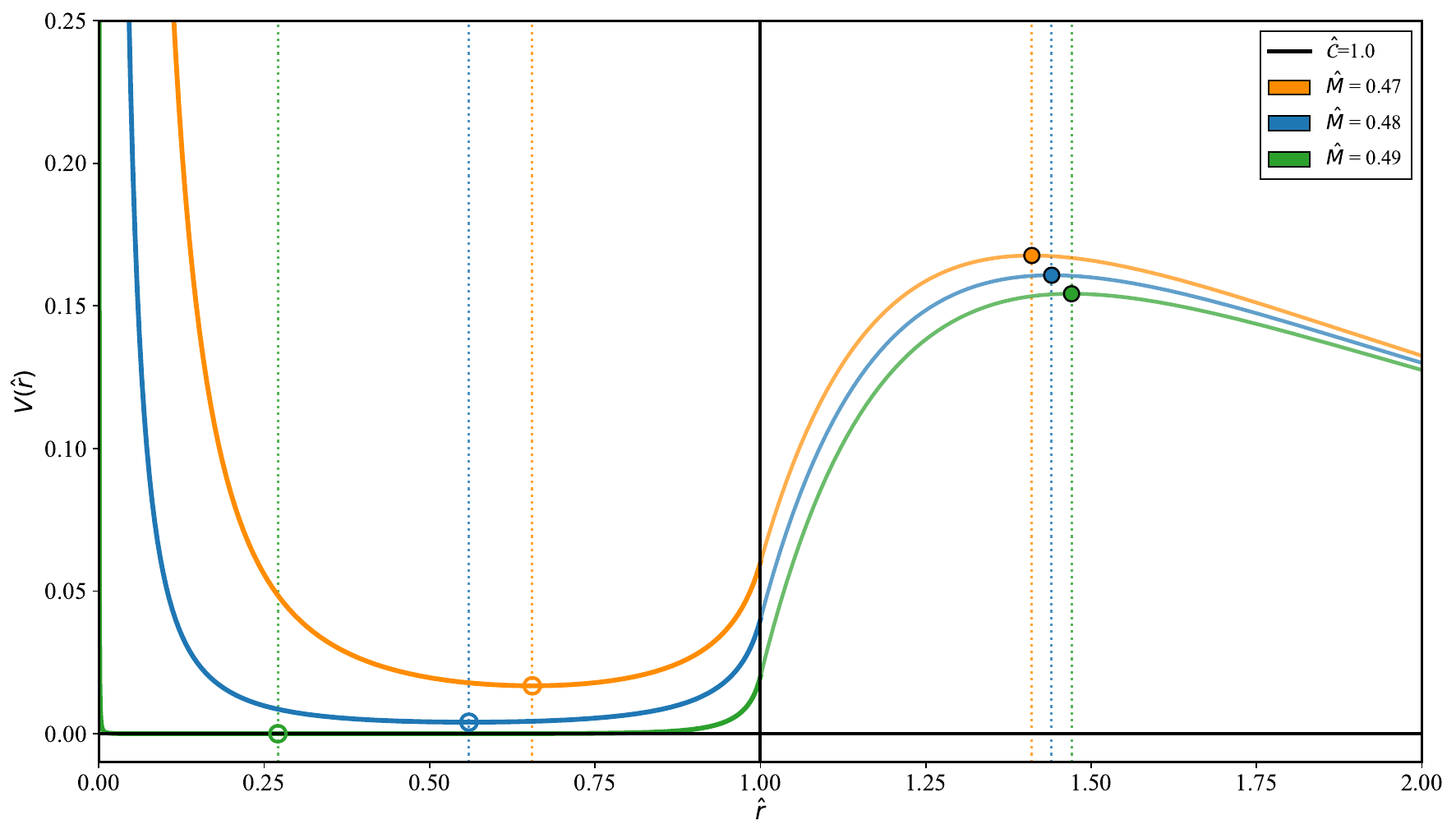}
    \caption{The figure shows the effective potential of  regular 
    anisotropic stars with $\hat{M} > \hat{M}_B$, including its critical points. There is a minimum in the interior of the object ($\circ$), and a maximum ($\bullet$) at the Schwarzschild value $\hat{r}_\gamma= 3\hat{M}$.
    }
\label{fig:effectivepotential}
\end{subfigure} 
    \hfill
\begin{subfigure}{0.5\textwidth}
    \centering
    \includegraphics[width=\linewidth]{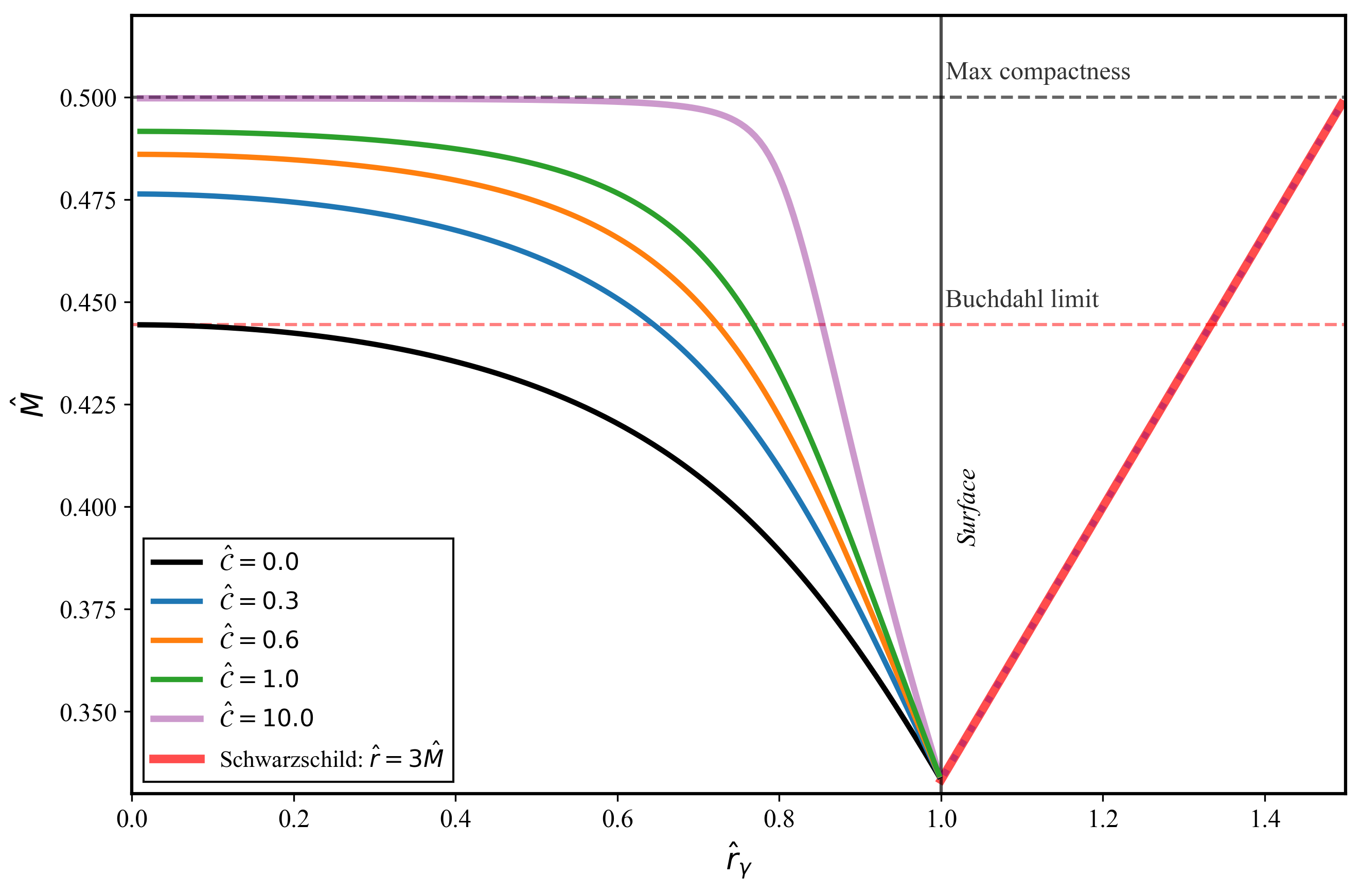}
    \caption{The figure shows the behaviour of the radius of the null circular orbits $\hat r_\gamma$ for increasing values of  $\hat M$ and different values of the anisotropic parameter $\ac$. The solid vertical line in black shows the surface of the object. For $\hat{M}\leq 1/3$, there exist no null circular orbits.
   }
\label{fig:photonring}
\end{subfigure}   
    \hfill
\begin{subfigure}{0.5\textwidth}
    \centering
    \includegraphics[width=\linewidth]{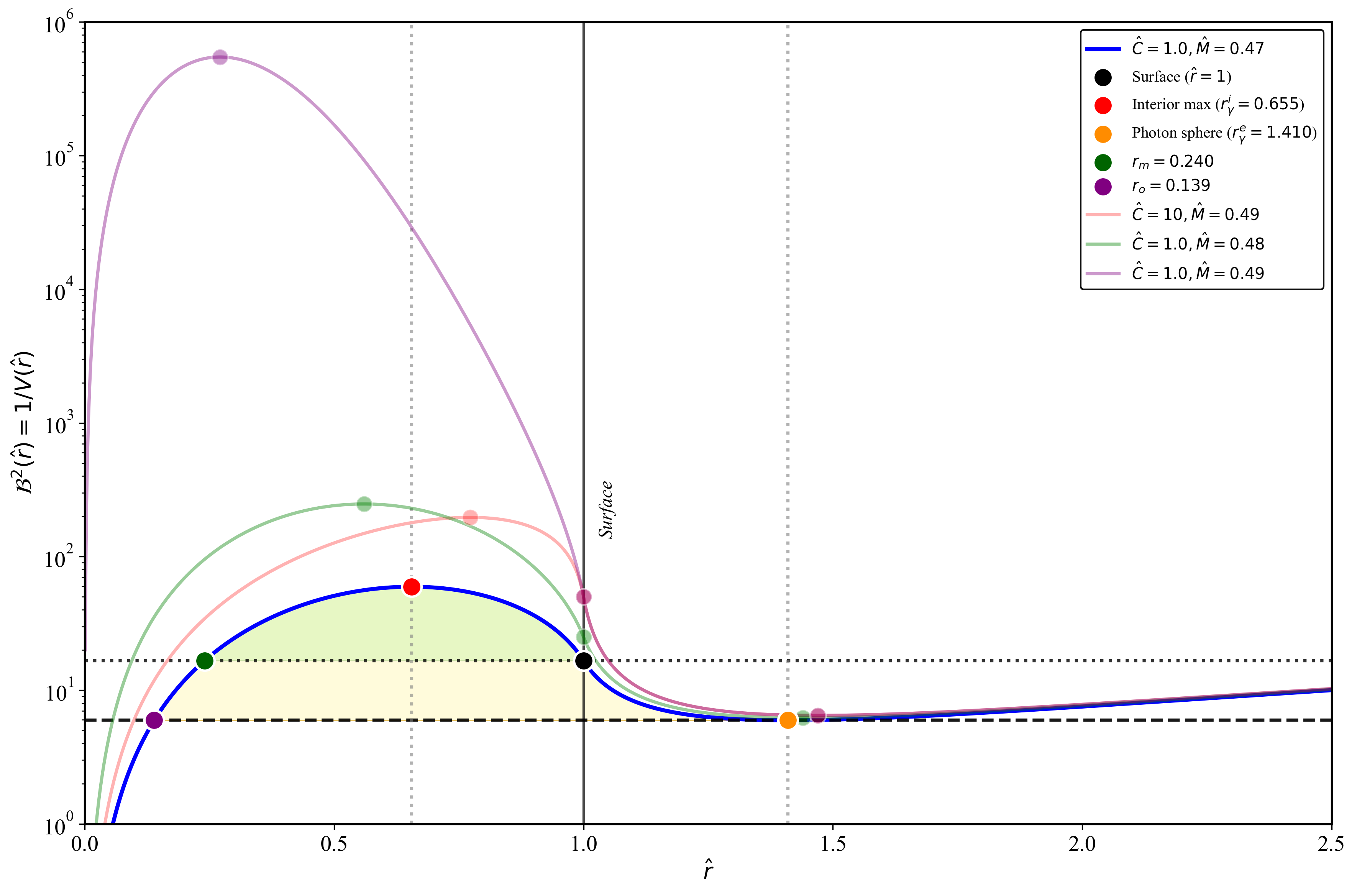}
    \caption{The figure  shows $\mathcal{B}^2(\hat{r})$ for massless particle trajectories in and around a compact object for several values of $\hat{M}$ and $\hat{C}$. For the curve in blue, particles with $\mathcal{B}^2$ in the region shaded in green remain trapped inside the object, while particles with $\mathcal{B}^2$ in the yellow region may leave the object but will reenter it.
    }
    \label{fig:impactfactor}
\end{subfigure}
    \caption{Effective potential, photon sphere and trapped regions}
    \label{fig:profiles}
\end{figure}

Using Eq.\eqref{gtt}, the effective potential for regular anisotropic stellar configurations can be written as follows:
\begin{equation}
    V = \frac{(1 -2\hat{M})}{\hat{r}^2}\exp \left[4 \hat{M} \int_{\hat{r}}^{1} \frac{\hat{x} d\hat{x}}{(3F(\hat{x}) - 1) (1 - 2\hat{M} \hat{x}^2)} \right] \ .
\end{equation}
The condition for circular null geodesics with $\hat r=\hat r_\gamma$ then reduces to,
\begin{equation}
    1 - 3 (1 - 2\hat{M}\hat{r}^2_\gamma)F(\hat{r_\gamma}) = 0\ .
\end{equation}
A key feature of very compact configurations is the emergence of an additional inner photon orbit at $\hat{r}_\gamma^{\rm in}$, located within the star \cite{Cardoso2014}, along with the unstable exterior orbit at $\hat r_\gamma^{\rm out}=3\hat M$ as depicted in Fig.\ref{fig:effectivepotential}. The former represents a stable circular null geodesic, whose position is sensitive to the anisotropy parameter. The effective potential for the isotropic case is shown in Appendix \eqref{app:iso}. For fixed compactness, an increment in the anisotropy shifts the orbit to larger radii relative to the isotropic case. As the compactness increases for fixed $\mathcal{\hat C}$, however, $\hat{r}_\gamma^{\rm in}$ moves inward toward the center, as illustrated in Figs.\ref{fig:photonring} and \ref{fig:impactfactor}. The figure also shows that a given value of $\hat{r}_\gamma^{\rm in}$ can be attained for different pairs of $(\hat M, \mathcal{\hat C})$. From Fig.\ref{fig:impactfactor} we also see that for a fixed $\hat M$ the depth of the potential decreases for increasing $\ac$. The exterior spacetime is that of the Schwarzschild geometry, with the outer circular null geodesic located at $\hat{r}_\gamma^{\rm out} = 3\hat{M}$ independently of the interior structure. Consequently, observations of the outer photon ring alone cannot distinguish between ultra-compact isotropic stars, anisotropic stars, or black holes when their Schwarzschild mass is the same.
Figure \ref{fig:impactfactor} shows $\mathcal{B}^2(r)\equiv 1/V(r)$. For values of $\mathcal{B}$ near the threshold $[\mathcal{B}^2(r_m), \mathcal{B}^2(r_o)]$, corresponding to the shallower region of the potential (yellow region), trajectories temporarily exit the star but are ultimately reflected back. There are also bound trajectories that cannot leave the star (green region). Null geodesics originating from infinity cannot access the trapped region, as they are separated by the potential barrier associated with the outer unstable photon sphere. The plot also shows that for higher $\ac$ and $\hat M$ the potential is shallower, which may indicate that such objects are more stable.

\section{Anisotropic Uniform Gravastars}
\label{gravastars}

We first summarize how a gravastar configuration follows from the isotropic incompressible star, in which the radial pressure is given by Eq.\eqref{isotropicPr}. As the compactness approaches the Buchdahl's critical value $\hat{M}_B = 4/9$, the central pressure required for hydrostatic equilibrium increases without bound. Precisely at $\hat{M} = \hat{M}_B = 4/9$, the divergence radius 
\begin{equation}
    \hat{r}_d = 3\sqrt{1 - \frac{4}{9\hat{M}}}\ ,
\end{equation}
coincides with the center, rendering the central pressure infinite. 

A characteristic property of the isotropic solution for $\hat{M} \geq \hat{M}_B$ is that the pressure takes a negative value at the center and diverges at a finite radius $0 \leq \hat r_d \leq 1$. As the compactness approaches the limiting value $\hat{M} = 1/2$, $\hat{r}_d$ migrates toward the stellar surface ($\hat{r}_d \to 1$), resulting in an interior dominated by a constant negative pressure $\hat P \sim -\hat{\rho}_*$, corresponding to a de Sitter-like core \cite{Mazur2015}.

\begin{figure}
    \centering
    \includegraphics[width=0.9\linewidth]{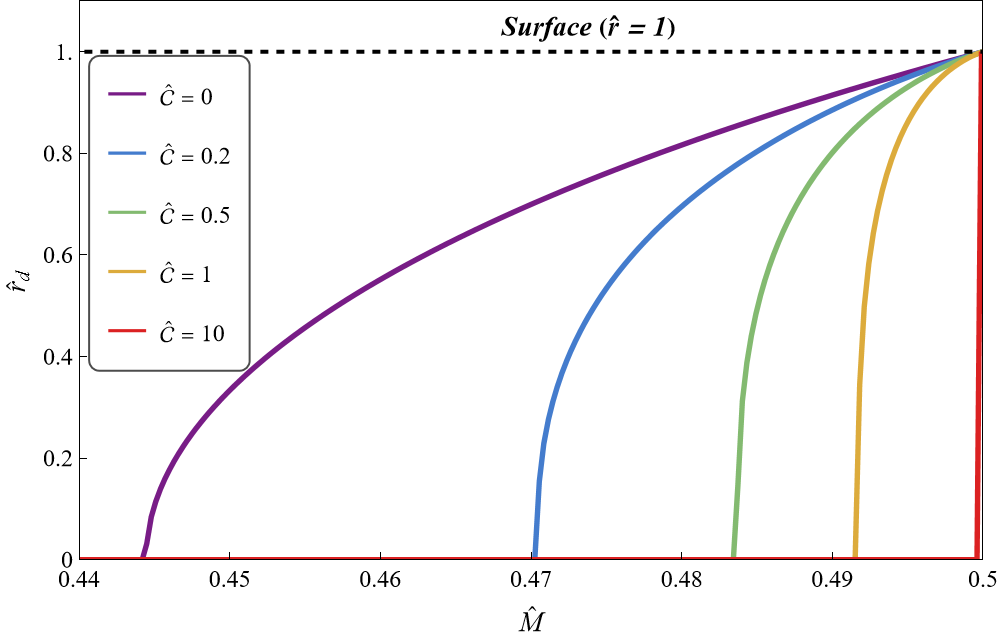}
    \caption{The figure shows the radial coordinate for which the pressures diverge as a function of the compactness, for different values of $\mathcal{\hat C}$.}
\label{fig:divergencecompactness}
\end{figure}

\begin{figure*}
    \centering
    
    \begin{subfigure}{0.325\textwidth}
        \centering
        \includegraphics[width=\linewidth]{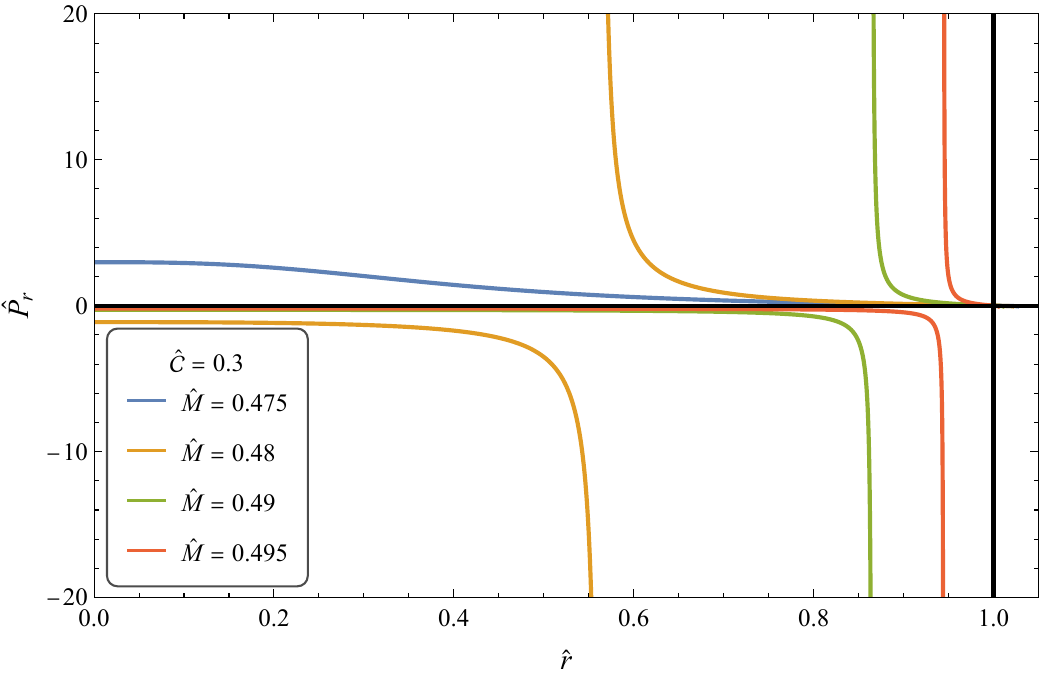}
        \caption{}
\label{fig:radial_pressure_grav}
    \end{subfigure}
    \hfill
    \begin{subfigure}{0.325\textwidth}
        \centering
        \includegraphics[width=\linewidth]{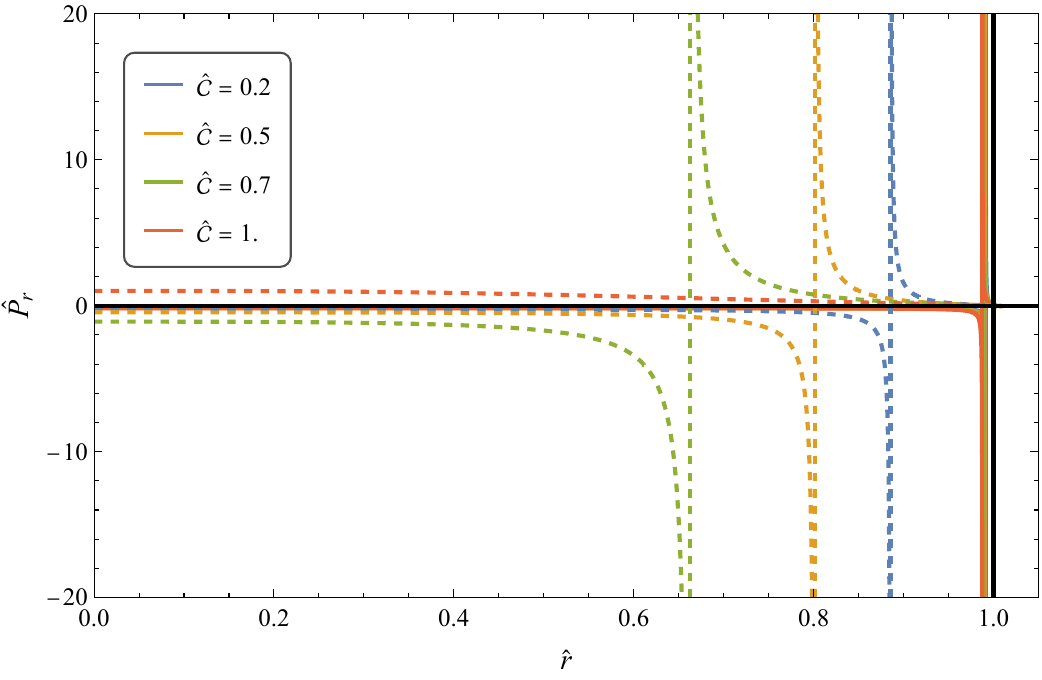}
        \caption{}
\label{fig:radialpressure1}
    \end{subfigure}
    \hfill
    \begin{subfigure}{0.325\textwidth}
        \centering
        \includegraphics[width=\linewidth]{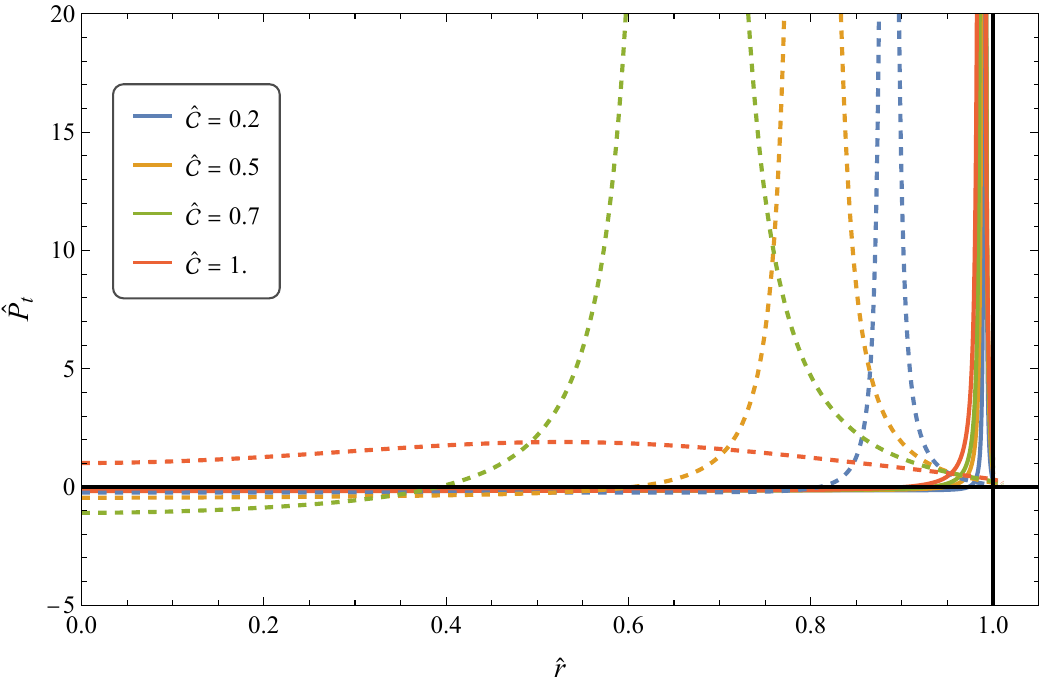}
        \caption{}
\label{fig:tangentialpress_grav}
    \end{subfigure}
    
    \caption{Profile of the radial pressure for (a) $\hat C=0.3$ and different values of $\hat M$, (b) $\hat{M} = 0.49$ (dashed line) and $\hat M=0.499$ (solid line) for different values of $\hat C$.
    Figure (c) shows the tangential pressure for $\hat{M} = 0.49$ (dashed line) and $\hat M=0.499$ (solid line) and several values of $\hat C$.}
\label{fig:profiles}
\end{figure*}

Anisotropic star configurations with $\hat{M}>\hat{M}_B$ display properties analogous to the isotropic case, with quantitative differences arising from the parameter $\mathcal{\hat C}$. The radial pressure is negative at the center and displays a first order pole at $\hat r=\hat r_d$ given by $1-3F(\hat M,\hat r_d,\ac)=0$ when $F<1/3$, see Eq.\eqref{radialpressure} and Fig.\ref{fig:energycondition}. The behaviour of $\hat r_d$ with $\hat M$ for several values of $\mathcal{\hat C}$ is shown in Figure \ref{fig:divergencecompactness}. For a fixed anisotropy parameter $\ac$, increasing the compactness shifts the divergence radius outward to the surface of the star. For fixed $\hat M$, the higher the value of $\ac$, smaller the divergence radius.
This behaviour is seen in Figs.\ref{fig:radial_pressure_grav} and \ref{fig:radialpressure1} (along with the tangential pressure profile in Fig.\ref{fig:tangentialpress_grav}), which show that both pressures for configurations with $\hat P_c \equiv \hat{P}_r(0)<0$ diverge at some $\hat r_d$. These configurations fall under the red (regular) region in Fig.\ref{fig:CMdivergecontour}.

\begin{figure}
    \centering
    \includegraphics[width=\linewidth]{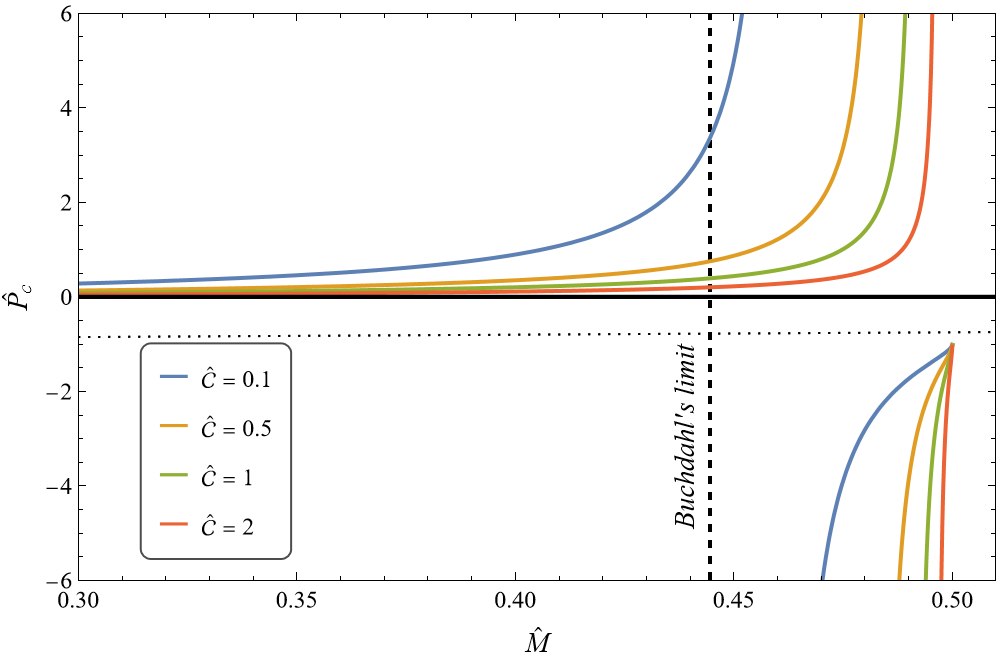}
    \caption{The plot shows the central pressure (in the units of $\hat{\rho}_*$) in terms of $\hat M$ for several values of $\ac$. $\hat{P}_c$ can be negative for highly compact anisotropic stars. Dotted line indicates $\hat{P}_c/\hat{\rho}_* = -1$.}
    \label{fig:pcaniso}
\end{figure}

\begin{figure}
    \centering
    \includegraphics[width=\linewidth]{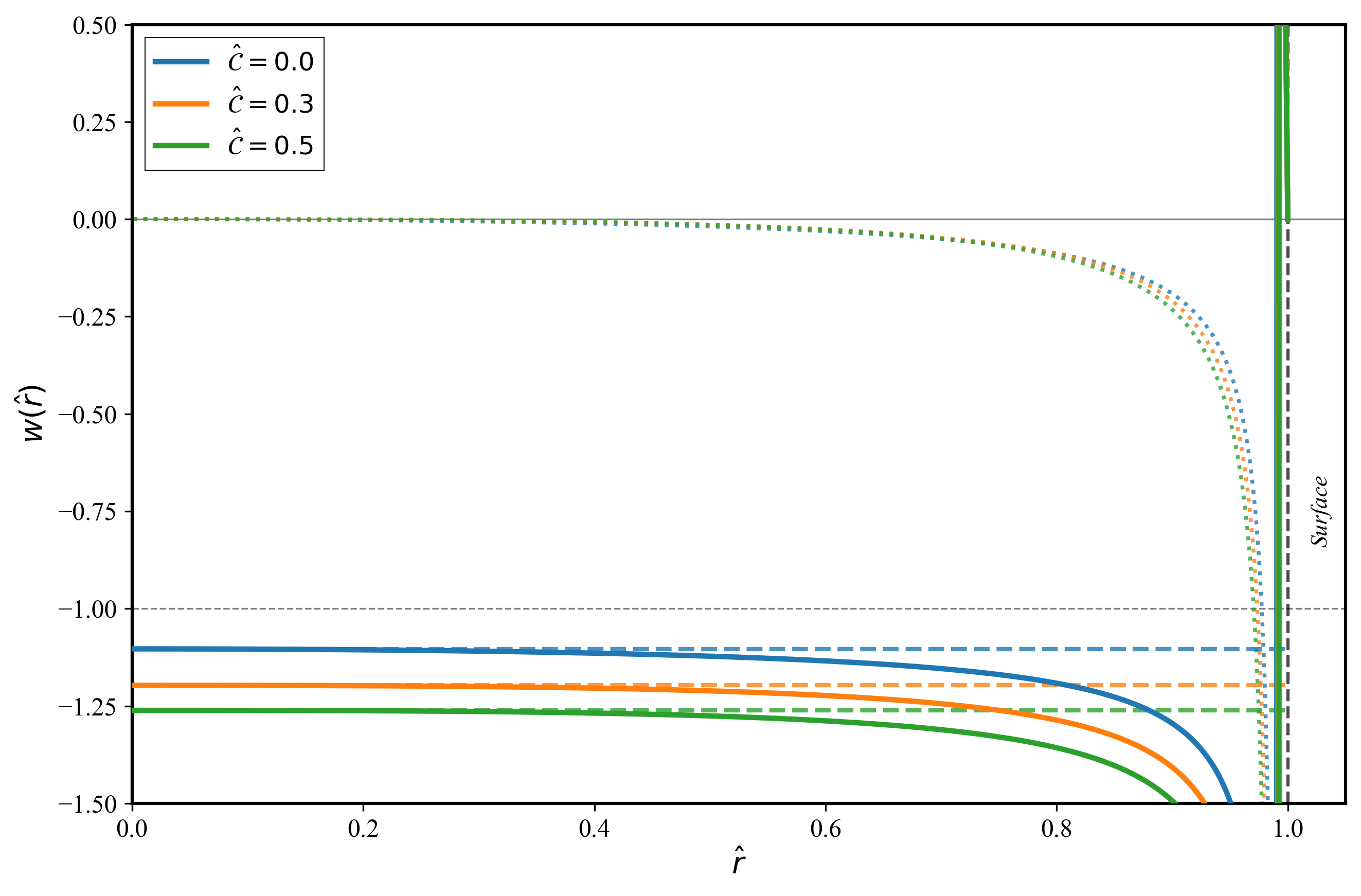}
    \caption{The figure shows $w = \hat{P}_r/\rho_*$ 
    as a function of $\hat r$ (thick lines). The dotted lines show the divergent part of $w$, while the dashed lines show the constant part. In all cases, $\hat{M} = 0.499$.}
    \label{fig:kds}
\end{figure}

Regarding the central pressure for the singular models, it increases (decreases) for fixed $\ac$ and increases $\hat M$ in the positive (negative) case; see Fig.\ref{fig:pcaniso}. For fixed compactness $\hat{M}$, increasing $\hat{\mathcal{C}}$ lowers the central pressure required for hydrostatic equilibrium.
For $\hat r\ll\hat{r}_d$ and high $\hat M$, the anisotropy $\Delta$ vanishes, and  the EOS is $\hat{P}_r \simeq -\kappa\hat\rho_*$, with a positive constant $\kappa$, see Fig.\ref{fig:kds}. This can also be seen analytically, since the radial pressure can be written as,

\begin{equation}
    \hat{P}_r(\hat r)=\hat{P}_c+
    \hat P_{\rm div}
\end{equation}
with 
\begin{equation}
    \hat{P}_c=\kappa\hat\rho_*\ ,\quad \hat P_{\text{div}}(\hat r)=\hat\rho_*\;[F(\hat r)-F(0)]\:\frac{1+3\hat{P}_c}{1-3F(\hat r)},
\end{equation}

and $ \kappa=\frac{F(0)-1}{1-3F(0)}\ $.

\begin{figure}
    \centering
    \includegraphics[width=0.95\linewidth]{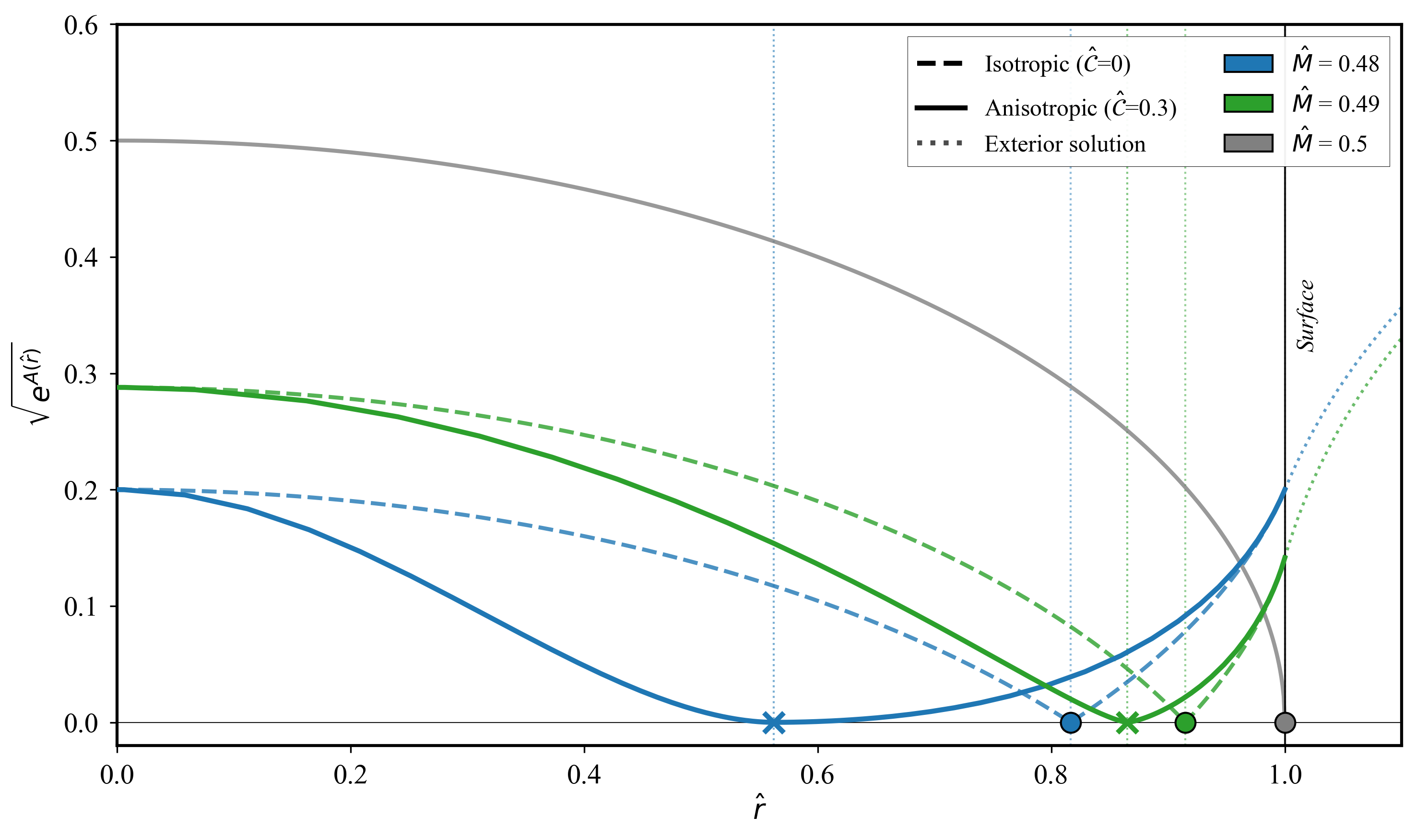}
    \caption{The plot shows $\sqrt{e^{A(\hat{r})}}$ for the singular configurations with and without anisotropic pressure in solid and dashed curves, respectively. The corresponding pressure divergence radii are indicated with $\times$ and $\bullet$.}
\label{fig:redshift}
\end{figure}

The plot of the redshift factor $\sqrt{e^A}$ is presented in Fig.\ref{fig:redshift}. The metric coefficient $g_{tt}$ has a zero at $\hat r=\hat{r}_d$ but remains positive everywhere. The zero approaches the Schwarzschild radius as $\hat{M}\to 1/2$.  In the exterior region, it coincides with the vacuum Schwarzschild's solution. 

The configurations with $\hat P_r(0)<0$ have three regions. The inner region $0<\hat{r}\ll \hat{r}_d$ becomes approximately isotropic toward the center, while the outer region $\hat{r}_d \ll \hat{r} < 1$ forms an anisotropic envelope that matches to the exterior Schwarzschild geometry at the surface. The intermediate anisotropic region contains the singular part of the pressure. As noted in \cite{Cattoen2005}, the singularity in the pressure implies a naked curvature singularity, since some of the orthonormal components of the Riemann tensor (and hence the Kretschmann scalar) diverge at $\hat r=\hat r_d$. As will be discussed next, the singular configurations can describe a gravastar after the replacement of the region containing the singularity by a thick shell.

\begin{figure*}
    \centering
    
    \begin{subfigure}{0.325\textwidth}
        \centering
        \includegraphics[width=\linewidth]{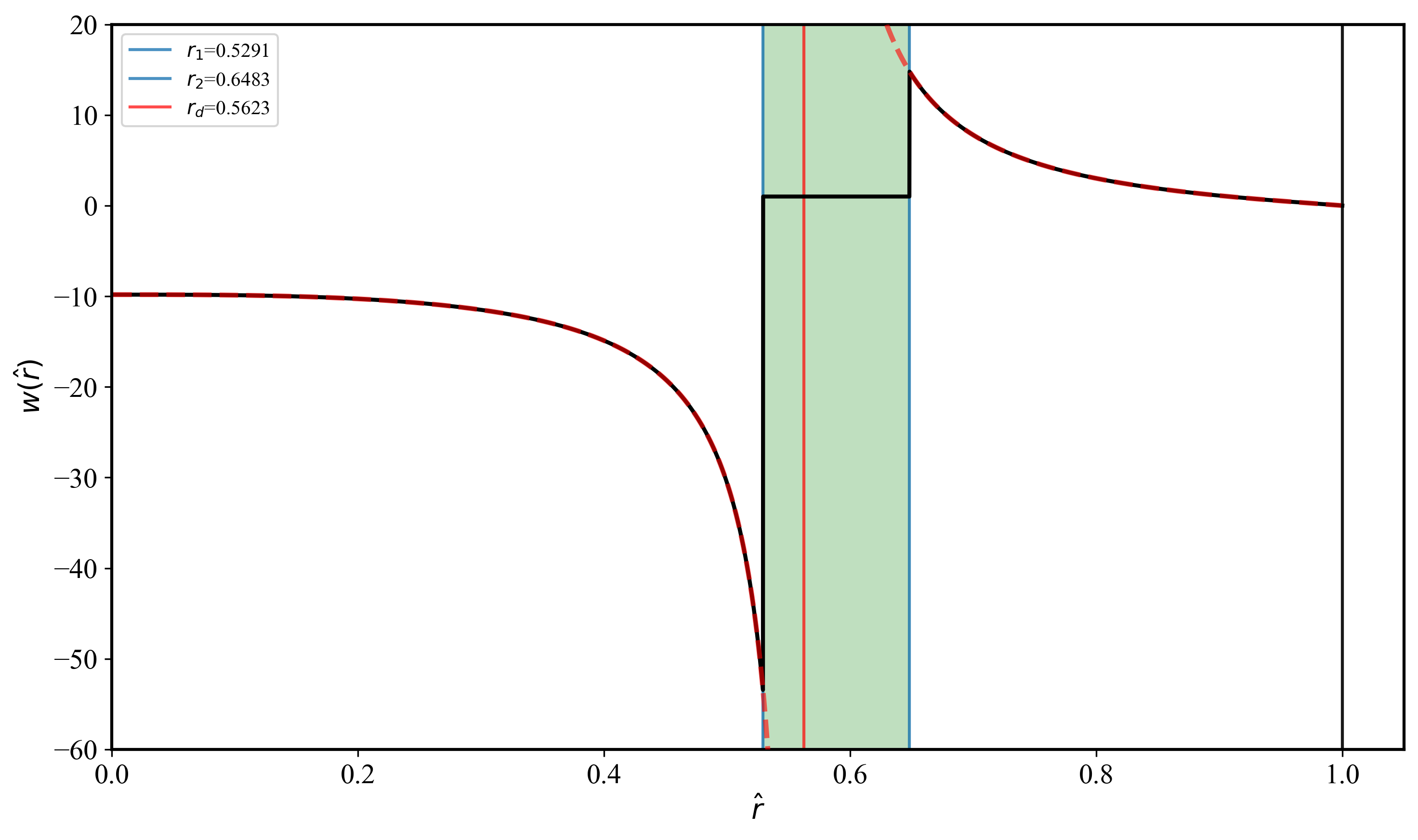}
        \caption{$\hat{M} = 0.48$}
        \label{fig:eosshell1}
    \end{subfigure}
    \hfill
    \begin{subfigure}{0.325\textwidth}
        \centering
        \includegraphics[width=\linewidth]{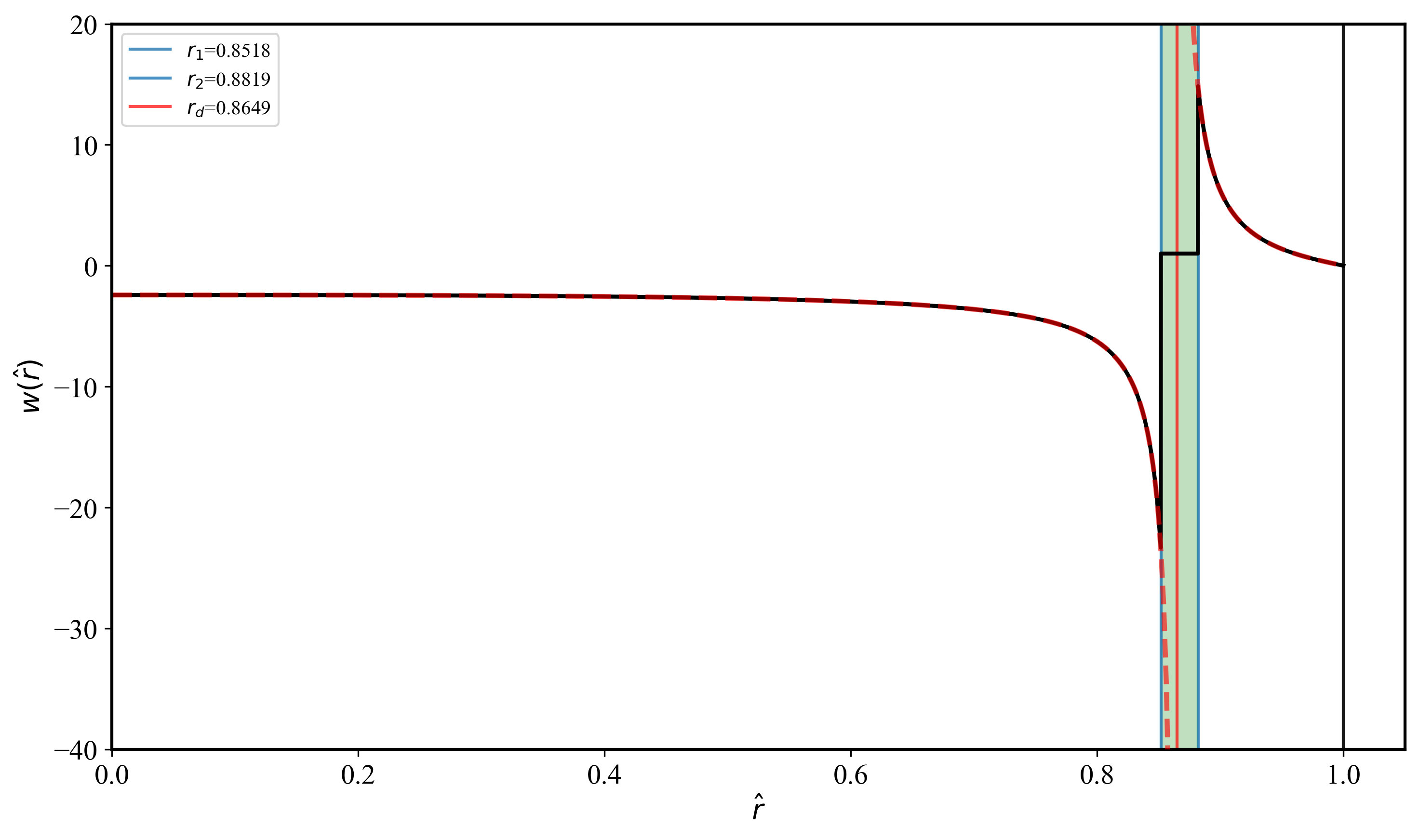}
        \caption{$\hat{M} = 0.49$}
        \label{fig:eosshell2}
    \end{subfigure}
    \hfill
    \begin{subfigure}{0.325\textwidth}
        \centering
        \includegraphics[width=\linewidth]{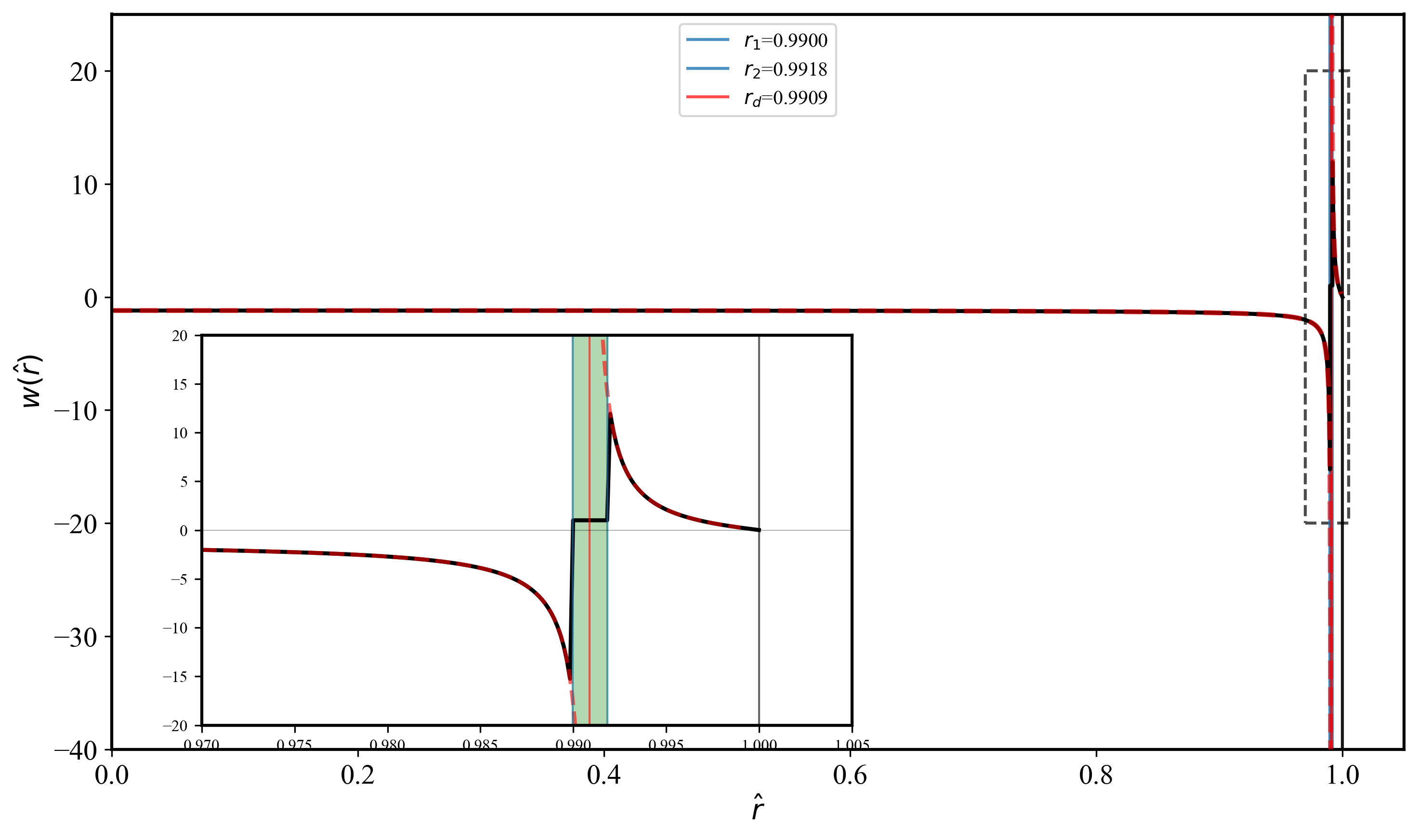}
        \caption{$\hat{M} = 0.499$}
        \label{fig:eosshell3}
    \end{subfigure}
    
    \caption{Profiles of the radial normalized pressure profile $w=\hat P_r/\hat\rho_*$ of a gravastar configuration, with $\hat{\mathcal{C}}=0.3$ and $\gamma=-2$ (stiff matter). Region II is depicted in green, and the vertical line shows where the divergence of the pressure would be.}
    \label{fig:profiles1}
\end{figure*}

\begin{figure*}
    \centering
    
    \begin{subfigure}{0.325\textwidth}
        \centering
        \includegraphics[width=\linewidth]{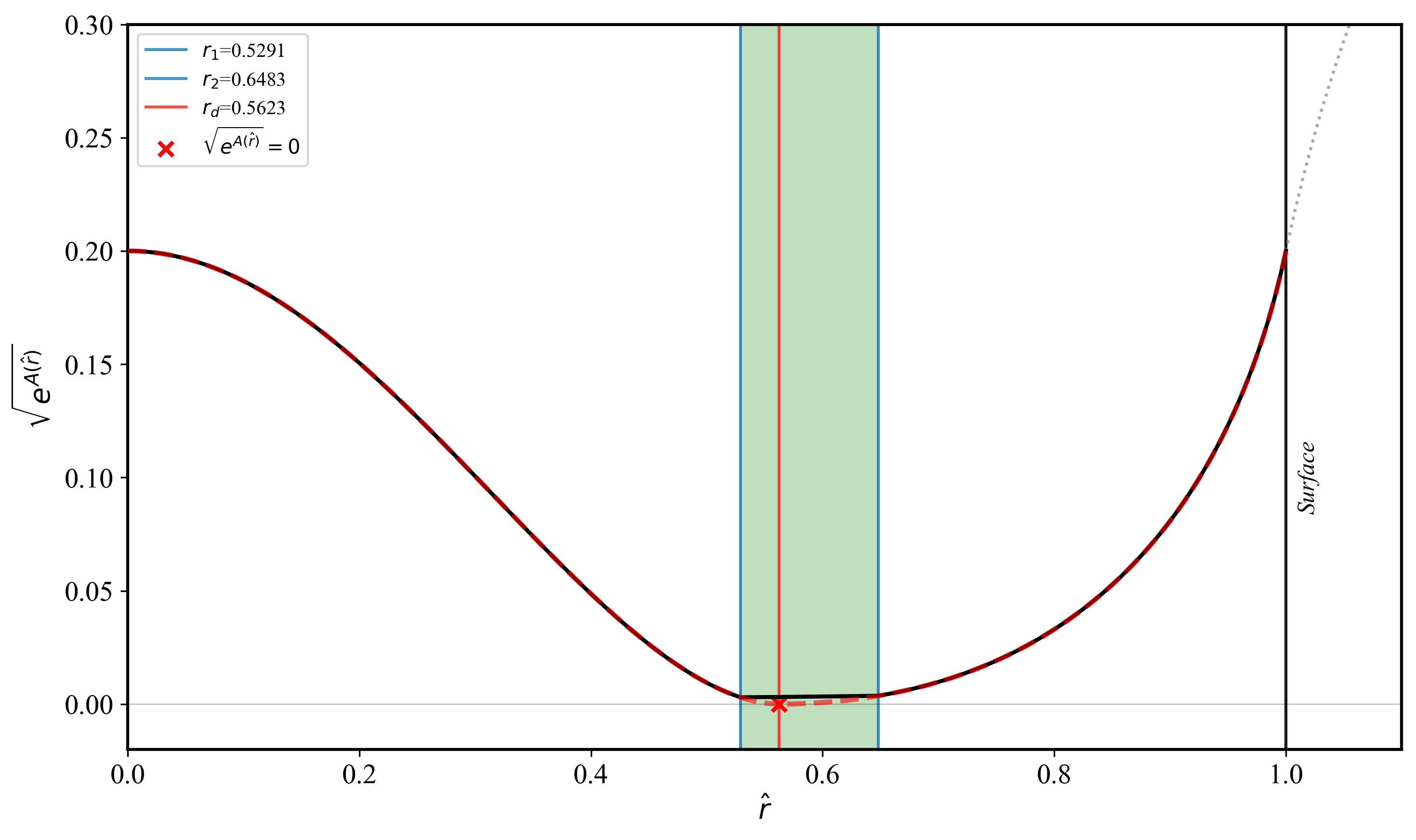}
        \caption{$\hat{M} = 0.48$}
        \label{fig:metricshell1}
    \end{subfigure}
    \hfill
    \begin{subfigure}{0.325\textwidth}
        \centering
        \includegraphics[width=\linewidth]{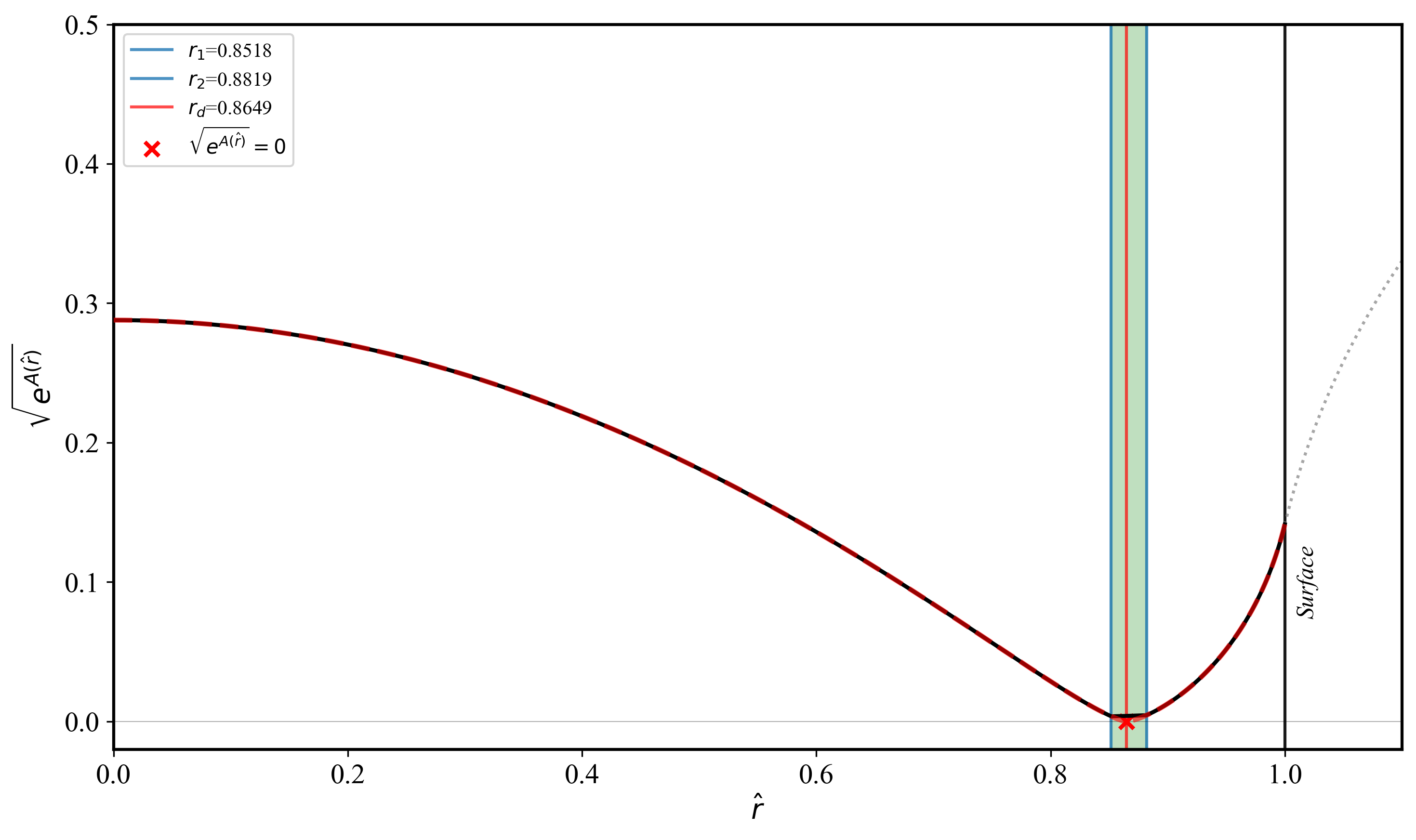}
        \caption{$\hat{M} = 0.49$}
        \label{fig:metricshell2}
    \end{subfigure}
    \hfill
    \begin{subfigure}{0.325\textwidth}
        \centering
        \includegraphics[width=\linewidth]{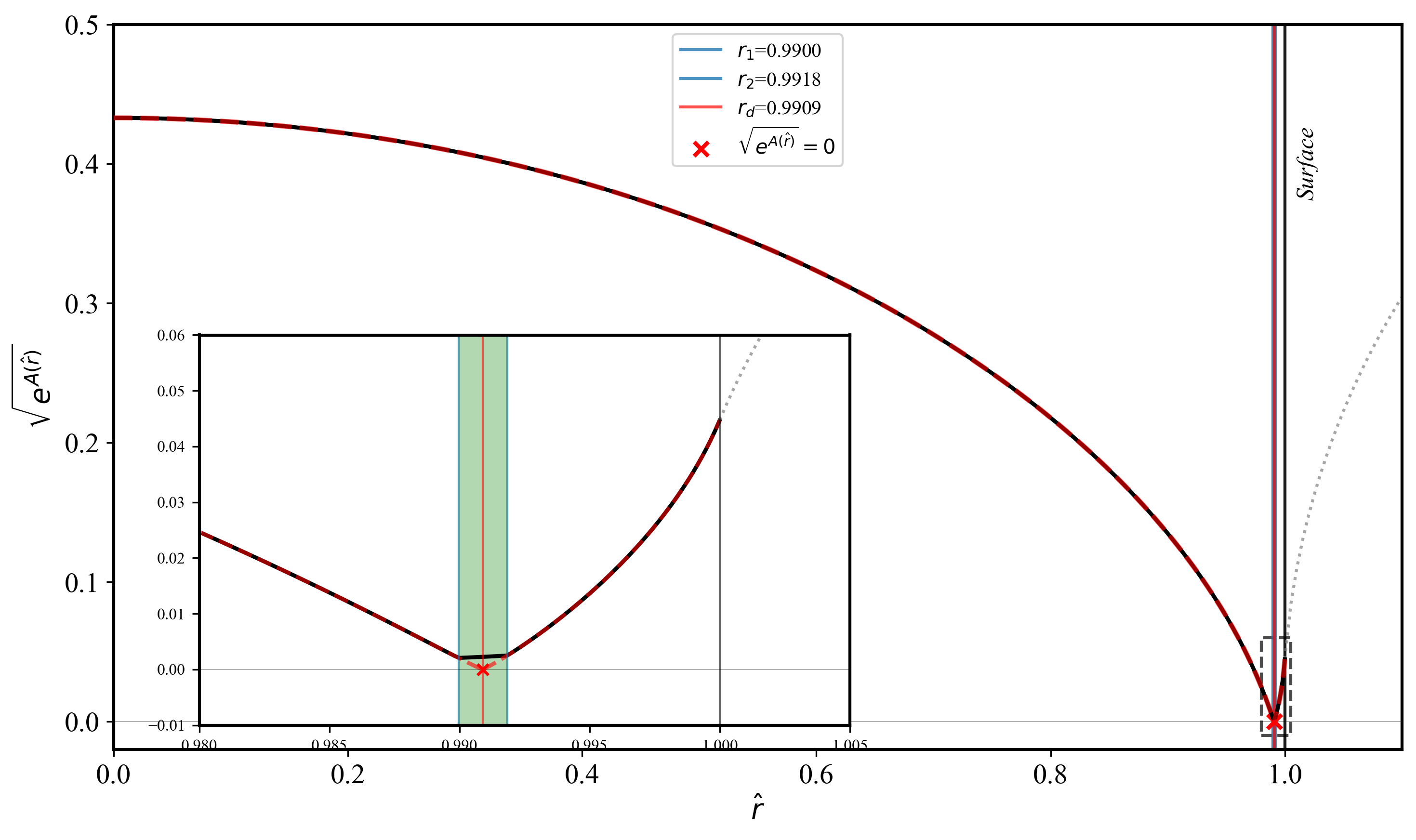}
        \caption{$\hat{M} = 0.499$}
        \label{fig:metricshell3}
    \end{subfigure}
    
    \caption{Redshift factor $\sqrt{g_{tt}}$ as a function of $\hat r$, for different values of $\hat M$, with $\hat{\mathcal{C}}=0.3$ and $\gamma=-2$.}
    \label{fig:profiles2}
\end{figure*}

\subsection{Thick shell}
\label{thickshell}

The main features of a gravastar configuration \cite{Cattoen2005} are the following: (\textit{i}) inside the gravastar, the density is everywhere positive and finite, (\textit{ii}) the central pressure is negative, (\textit{iii}) there are no event horizons in the associated spacetime. The original model presented in \cite{Mazur2015, Mazur2001} satisfies $P(0)=-\rho(0)$. As we shall see below, the anisotropic configurations presented here do not necessarily obey this condition. 

In order to have a regular configuration describing a gravastar, the region around the divergence will be replaced with a thick shell composed of a certain type of matter, which will be joined, using the junction conditions, to the anisotropic solution on both sides. Here we shall restrict ourselves to a simple realization of this model to show its more relevant features. We shall assume that the matter in the shell has the same constant density as the anisotropic regions, namely $\hat\rho_* = 3\hat M/4\pi$, so that $e^{-B} = 1 - 2\hat M\hat r^2$ everywhere inside the star. We shall also assume that the radial pressure in the shell has a constant value $\hat P_0$. Then the equation for the dimensionless radial pressure becomes

\begin{equation}
    8\pi \hat P_0 = \frac{(1 - 2\hat M\hat r^2)}{\hat r}\frac{dA}{d\hat r} - 2\hat M ,
\end{equation}

and it follows that

\begin{equation}
    \frac{dA}{d\hat r} = -\frac{4\gamma \hat M\hat r}{1 - 2\hat M\hat r^2} ,\quad {\rm where}\:\: \gamma \equiv -\frac{1}{2}\left(\frac{4\pi \hat P_0 + \hat M}{\hat M}\right).
\end{equation}

Hence, 

\begin{equation}
    \hat P_r (\hat r) = \hat P_0 = -\frac{\hat\rho_*}{3}(2\gamma + 1).
\end{equation}

This solution is valid for any value of $\gamma$. For $\gamma = 1$ this yields the vacuum EOS, while $\gamma = -2$ corresponds to the stiff matter EOS (for example, used in \cite{Mazur2015}).

The equation for $A(\hat r)$ is easily integrable, and it yields

\begin{equation}
    e^{A_{\mbox{\tiny II}}} = \mathcal{K}_{\mbox{\tiny II}} (1 - 2\hat M\hat r^2)^\gamma,
\end{equation}
where $\mathcal{K}_{\mbox{\tiny II}}$ is an integration constant, and the subindex $\mbox{II}$ denotes the central region of the configuration. The tangential pressure
is given by the third equation in \eqref{toveqs}:
\begin{equation}
    \hat P_{t_{\mbox{\tiny II}}}(r) = \frac{  2\hat M^2\hat r^2(\gamma^2 + \gamma + 1)-\hat M(2\gamma+ 1)}{4\pi(1 - 2\hat M\hat r^2)}\ .
\end{equation}
Hence, the anisotropic pressure $\Delta = P_{t{\mbox{\tiny II}}} - P_0$ is

\begin{equation}\label{aniso_slab}
    \Delta = \frac{\gamma (\gamma -1)\hat M^2 \hat r^2}{2\pi (1 - 2\hat M \hat r^2)} \ .
\end{equation}

The configuration that results from the introduction of a thick shell through the application of junction conditions to regularize the divergence consists of three regions.\footnote{$e^{-B} = 1 - 2\hat M\hat r^2$ in all three regions. Hence, the presence of the shell does not alter the Misner-Sharp mass of the object.}
:
\begin{enumerate}
    \item Region I, defined by $0<\hat r<\hat r_1$, is described by the anisotropic solution, namely, 
    \begin{equation}
        A_{\mbox{\tiny I}}(\hat{r})  = 4 \hat{M} \int^{\hat{r}}_{\hat{r}_1} \frac{\hat{r}' d\hat{r}'}{(3F(\hat{r}') - 1) (1 - 2\hat{M} \hat{r}'^2)} +{k}_{\mbox{\tiny I}}\ ,
    \end{equation}
    \begin{equation}
        \hat P_{r{\mbox{\tiny I}}} = \hat\rho_*\frac{F(r)-1}{1-3F(r)}\ ,
    \end{equation}
    where $ {k}_{\mbox{\tiny I}}$ is an integration constant to be fixed by the boundary condition at the center.
    \item The constant density and constant radial pressure solution is adopted in region II ($\hat r_1<\hat r<\hat r_2$), 
    namely
    \begin{equation}
        e^{A_{\mbox{\tiny II}}} = \mathcal{K}_{\mbox{\tiny II}}(1-2\hat M\hat r^2)^\gamma,
    \end{equation}
    \begin{equation}
        P_{r{\mbox{\tiny II}}} = -\frac{\hat\rho_*}{3}(2\gamma+1),
    \end{equation}
    \begin{equation}
        \hat P_{t{\mbox{\tiny II}}} = \frac{-\hat M(2\gamma+1) + 2\hat M^2\hat r^2(\gamma^2+\gamma+1)}{4\pi(1-2\hat M\hat r^2)}\ .
    \end{equation}
    \item In region III, defined by $\hat r_2<\hat r<1$, we adopt the anisotropic solution, as given in region I, with an integration constant ${k}_{\mbox{\tiny III}}$
     fixed by matching 
     $A_{\mbox{\tiny III}}$
     to Schwarzschild's solution at the surface.
\end{enumerate}
As a simplifying assumption, in the models described here we choose to fix $k_{\mbox{\tiny I}}$ in such a way that $A_{\mbox{\tiny I}}$ takes at $\hat r = 0$ the value of the isotropic model, given by Eq.\eqref{eA_ISO}.

We show in Fig.\ref{fig:profiles1} the radial pressure as a function of $\hat r$ for several values of the compactness, with   $\mathcal{\hat C}=0.3$, $\gamma=-2$, and region II indicated in green. The redshift factor is shown in  Fig.\ref{fig:profiles2}, with the same values of the parameters. These figures follow from the imposition of the junction conditions at $\hat r=\hat r_1$ and $\hat r=\hat r_2$, with $\gamma=-2$, see Appendix \eqref{jc} for details.

A distinctive feature of these configurations is that, as $\hat M$ increases, the shell moves closer to the surface and its thickness decreases. As the shell moves to the surface, most of the interior of the configuration tends to the de Sitter spacetime. Another feature is that the NEC is violated in region I. Such a violation, present also in the isotropic case (see below), is unavoidable, but the region where it occurs could be reduced either by considering a thicker shell or high compactness. In the latter case, the NEC violation is restricted to a region very near the shell (see Fig.\ref{fig:eosshell3}).

The DEC is violated in regions II and III (and in region I as well, due to the violation of the NEC), but the size of regions II and III in this model is reduced for large $\hat M$. The jump in the pressure from region I to region II decreases with increasing $\hat M$ as well, minimizing DEC violation.

\begin{figure}
    \centering
\begin{subfigure}{0.5\textwidth}
    \centering
    \includegraphics[width=\linewidth]{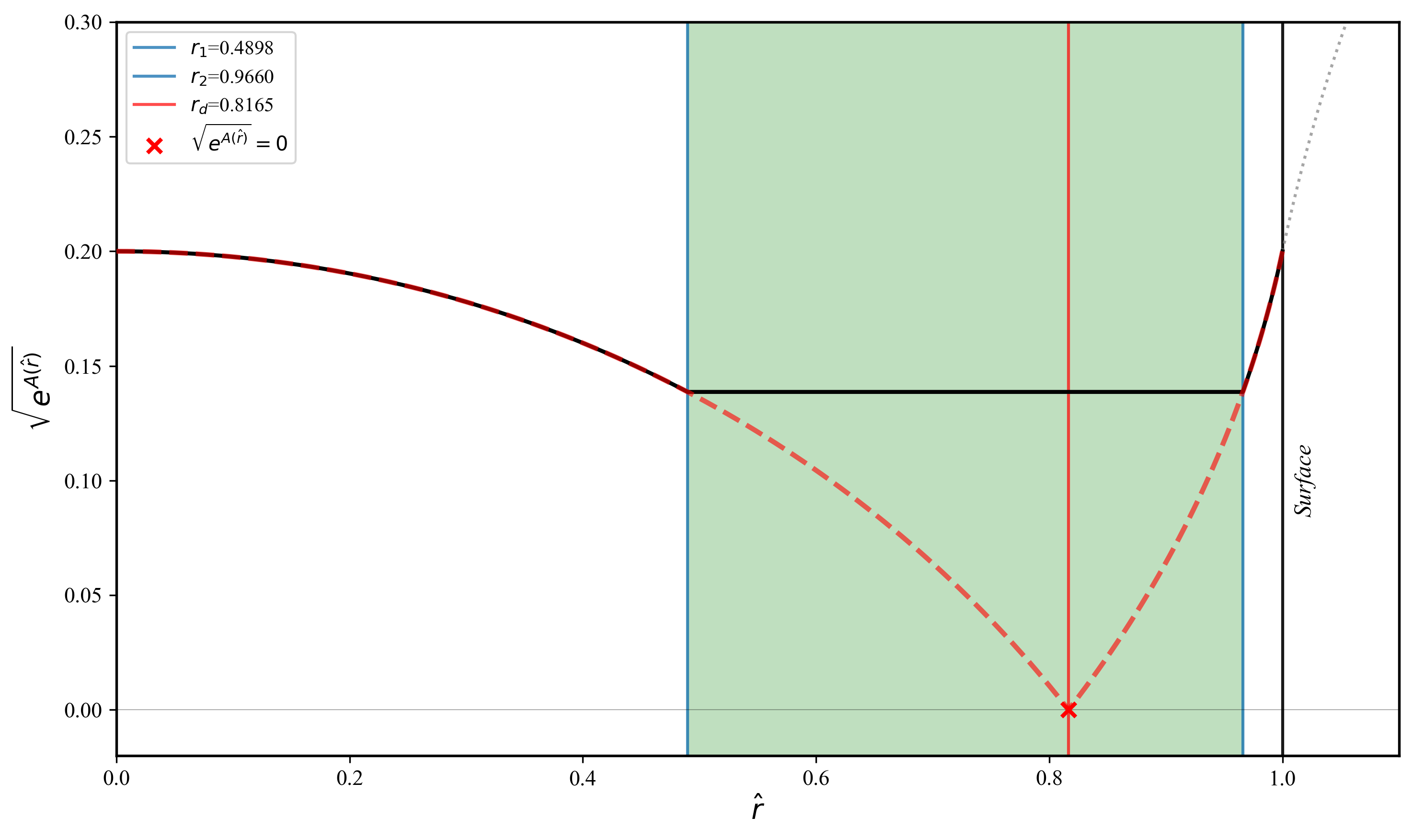}
    \caption{$\hat{M} = 0.48$.
   }
\label{fig:isometric48}
\end{subfigure}   
    \hfill
\begin{subfigure}{0.5\textwidth}
    \centering
    \includegraphics[width=\linewidth]{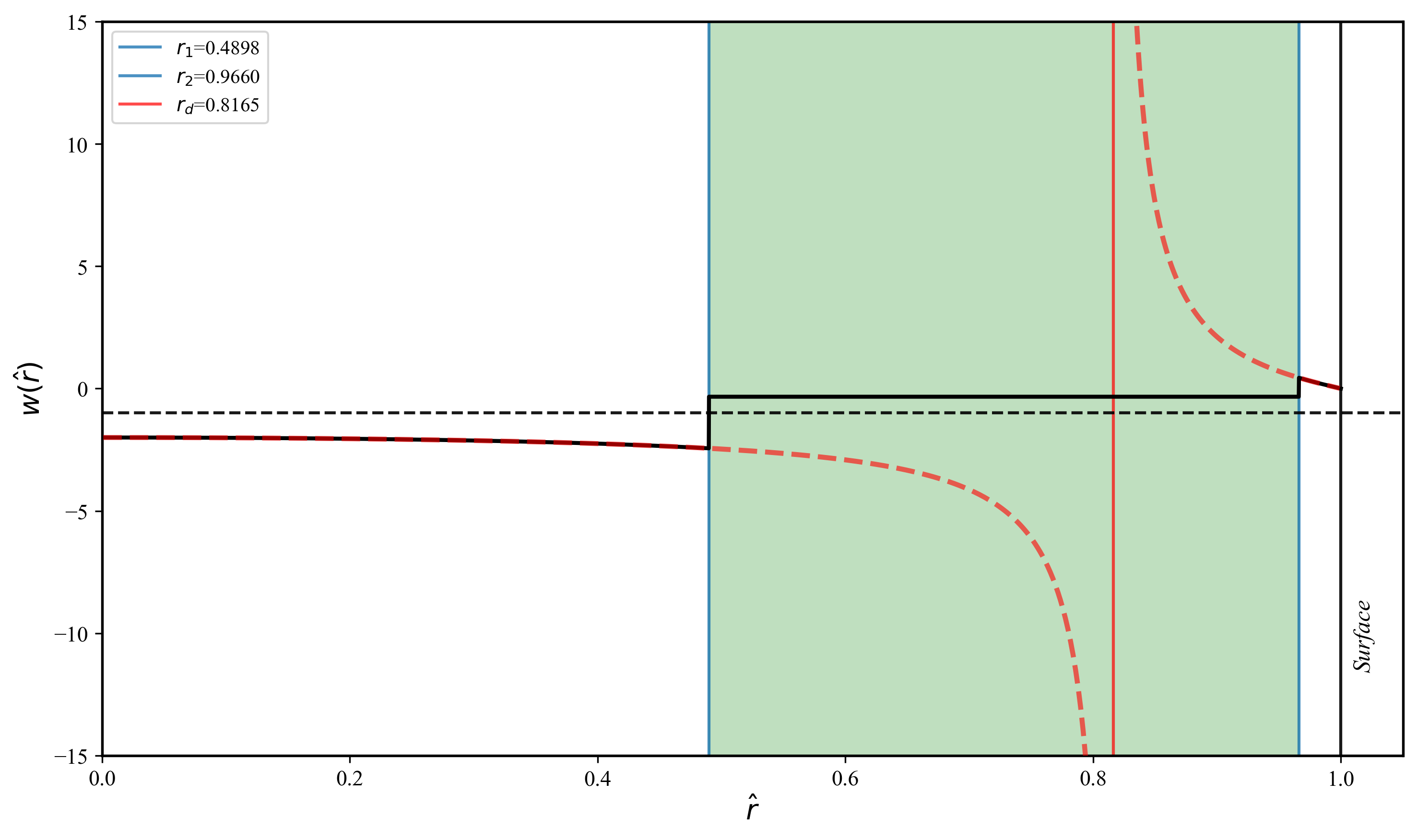}
    \caption{$\hat{M} = 0.48$}
    \label{fig:isoeos48}
\end{subfigure}
    \caption{Redshift factor and normalized pressure profile for the fully isotropic gravastar with $\gamma=0$ and $\hat{M}=0.48$. The dashed line corresponds to $w=-1$}
    \label{fig:isoprofiles48}
\end{figure}

\begin{figure}
    \centering
\begin{subfigure}{0.5\textwidth}
    \centering
    \includegraphics[width=\linewidth]{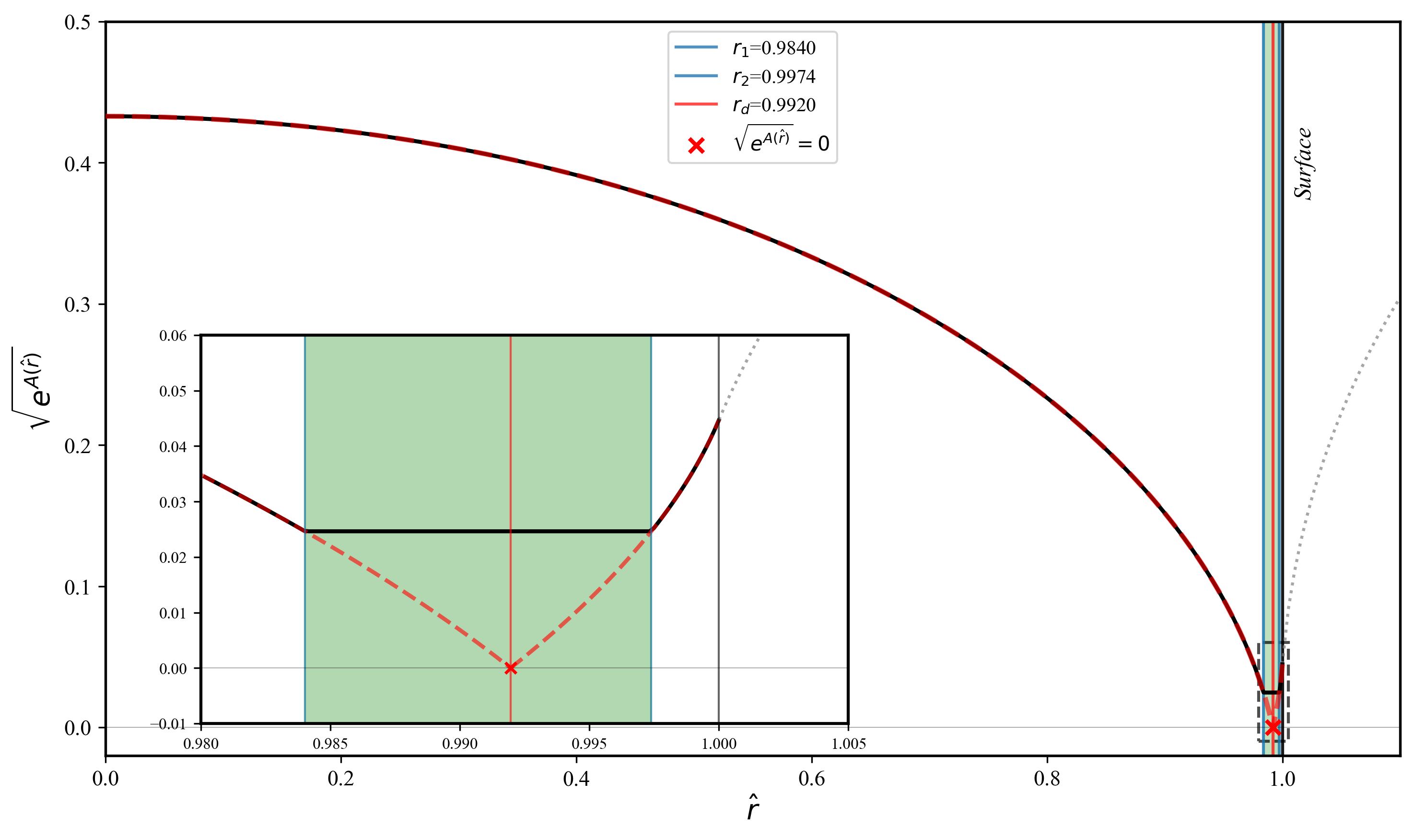}
    \caption{$\hat{M} = 0.499$    }
\label{fig:isometric499}
\end{subfigure}   
    \hfill
\begin{subfigure}{0.5\textwidth}
    \centering
    \includegraphics[width=\linewidth]{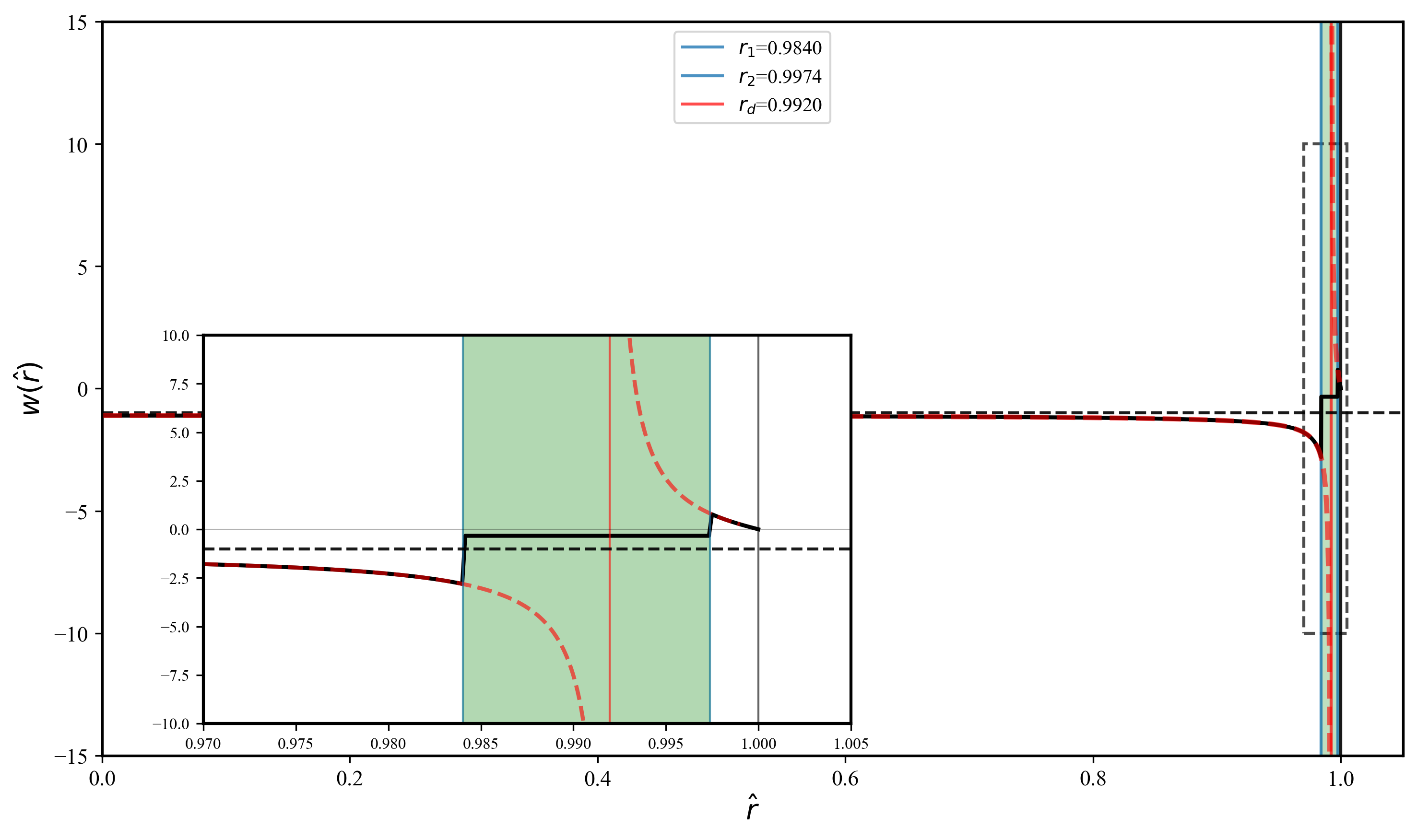}
    \caption{$\hat{M} = 0.499$ }
    \label{fig:isoeos499}
\end{subfigure}
        \caption{Same as Fig.\ref{fig:isoprofiles48}, but for the higher compactness $\hat{M}=0.499$.}
    \label{fig:isoprofiles499}
\end{figure}

Let us point out that a completely isotropic gravastar can be obtained as a particular realization of the model presented above, namely in the limit $\ac=0$. As an example, we present the plots corresponding to $\gamma=0$, see Figs.\ref{fig:isoprofiles48} and \ref{fig:isoprofiles499}. This isotropic configuration does not contradict the findings in \cite{Cattoen2005}, since the isotropic model presented here violates the NEC in region I, and does not display a continuous pressure \footnote{The violation of the NEC is also inherent to the homogeneous isotropic configurations beyond the Buchdahl limit, see for instance \cite{Posada2018}}. The qualitative features of the isotropic configurations coincide with those of the anisotropic configurations, but both the NEC and the DEC violations are weaker. 

\section{Final remarks}\label{concl}

We have presented the details of the anisotropic configurations that follow from considering a constant density together with the covariant EOS introduced in \cite{Raposo2018}. The resulting model can be treated analytically, except for the metric coefficient $g_{tt}$, given in terms of an integral to be solved numerically. 

Among the various types of objects described by the model, those attaining high compactness are perhaps the most interesting. We have shown that such compactness can be achieved in two distinct ways. For sufficiently large values of the anisotropy parameter $\ac$, objects with positive central pressure can attain compactness arbitrarily close to the black-hole limit. In these configurations, the tangential pressure is negative in the interior but increases and becomes positive near the surface, leading to a localized violation of the DEC.

The second possibility to attain high values of $\hat M$ is to consider objects with negative central pressure. In this case, the negative pressure can be regularized through the introduction of a thick shell, with the appropriate junction conditions, leading to the three-layer gravastar configurations considered here. These objects can reach values of $\hat M$ close to $0.5$ even for low values of $\ac$. We have also shown that isotropic gravastars are possible in this setting, without contradicting previous results \cite{Cattoen2005}, due to the violation of the NEC in region I and discontinuous pressures across the regions.

The use of these models as black hole mimickers ultimately depends on features such as their stability (\cite{Posada2018,Visser2003}), and the addition of rotation \cite{Beltracchi2025}. The zero-thickness shell limit of the gravastar configurations presented here and its generalizations may also be of interest, to investigate the possibility of obtaining a final state different from that of de Sitter inside the object. We hope to return to these issues in future publications.

\begin{acknowledgments}
PHP is financially supported by \textit{Fundação de Amparo à Pesquisa e Inovação do Espírito Santo} (FAPES, Brazil) and the Coordena\c c\~ao de Aperfei\c coamento de Pessoal de Nível Superior (CAPES, Brazil). PHP thanks Departamento de F\'isica Te\'orica (UERJ) for hospitality. SEPB acknowledges support from of CNPq, grant PQ-C 300241/2025-9.
Nora Bretón and SEPB would like to thank Florencia Anabella Teppa Pannia for her contributions during the early stages of this work.
\end{acknowledgments}

%

\appendix

\section{The isotropic case}
\label{app:iso}

In the isotropic limit, the metric function $e^{A(r)}$ takes a simple form in the interior of the star, namely,
\begin{equation}\label{eA_ISO}
    e^{A(\hat{r})}_{\rm iso} =  \frac{1}{4}\left( 3\sqrt{1 - 2\hat{M}} - \sqrt{1 - 2\hat{M}\hat{r}^2}  \right)^2 \ .
\end{equation}
It remains positive everywhere inside the star, and reaches zero at $\hat r=\hat{r}_d$. Now, this is also the radius where the isotropic pressure diverges. This configuration was studied in \cite{Mazur2015}, where it was shown that the divergence can be regularized with a thin shell which must display anisotropic stresses.

The effective potential in the isotropic case can be obtained in closed form. In the interior and exterior regions, we have,
\begin{equation}
V(\hat{r}) = \begin{cases}
    V_{\rm in} = \frac{1}{4\hat{r}^2} \left(3\sqrt{1 - 2\hat{M}} - \sqrt{1- 2\hat{M}\hat{r}^2}\right)^2\quad & \hat{r} \leq 1\\
    V_{\rm ext} = \frac{\hat{r} - 2\hat{M}}{\hat{r}^3} \quad & \hat{r} > 1
\end{cases}
\end{equation}

Extremizing $V_{\rm in}$ for $\hat r\leq 1$ yields two roots, 
\begin{equation}\label{extphotonring}
    \hat{r}_a = \sqrt{\frac{\hat{M}_B - \hat{M}}{\hat{M}(1 - 2\hat{M})}}\ , \quad \hat{r}_b = 3 \sqrt{1 - \frac{\hat{M}_B}{\hat{M}}}.
\end{equation}
Only those roots satisfying $0 < \hat{r} \leq 1$ correspond to physically admissible photon orbits. For $1/3 \leq \hat{M} \leq \hat{M}_B$, a single circular null geodesic (associated to $\hat r_1$) exists within the star. For stars with $\hat{M} \geq \hat{M}_B$, only $\hat{r}_b$ is relevant. It starts at the stellar surface and subsequently moves inward as $\hat{M} \to 1/2$. 

The presence of a local minimum in the effective potential leads to the existence of trapped null trajectories. Specifically, for $\mathcal{B}^2\in\left[\frac{1}{V_{\max}},\frac{1}{V_{\min}}\right]$, null geodesics are confined between two radial turning points and remain trapped within the stellar interior.

\section{Junction conditions}\label{jc}

Here we present a summary of the method that permits the joining of two spacetimes \cite{Israel1966}, as well as its application to the case at hand. If the two geometries, denoted here by "+" and "-" are joined along a timelike 3-surface $\Sigma$ with normal vector $n^\mu$, the following conditions must be satisfied  \cite{Visser1995}:

\begin{equation}\label{cont}
    g^+_{\mu\nu}|_\Sigma=g^-_{\mu\nu}|_\Sigma,
\end{equation}

\begin{equation}
    S_{\mu\nu}=-\frac{1}{8\pi}([K_{\mu\nu}]-[K] h_{\mu\nu}),
    \label{smunu}
\end{equation}

where $S_{\mu\nu}$ is the stress-energy tensor of the matter on $\Sigma$, $[A]\equiv A^+-A^-$ for any quantity $A$, $h_{\mu\nu}\equiv g_{\mu\nu}-n_\mu n_\nu$, and the second fundamental forms $K^{\pm}_{\mu\nu}$ are

\begin{equation}
    K^{\pm}_{\mu\nu}\equiv \frac{1}{2}(\nabla^\pm_{\mu}n_\nu+\nabla^\pm_{\nu}n_\mu).
\end{equation}

In summary, these expressions show that the geometries $g^\pm_{\mu\nu}$ determine the possible matter content on $\Sigma$.

We apply this formalism to the three-region configuration described in Sect.\eqref{thickshell}, separated by timelike surfaces at $\hat r=\hat r_1$ and $\hat r=\hat r_2$. Since $e^{-B}=1-2\hat M\hat r^2$ is shared by all three regions, $[g_{rr}]=0$ holds at both $\hat r_1$ and $\hat r_2$, as well as $K^{\pm}_{\theta\theta} =K^{\pm}_{\phi\phi}=0$. The only quantity that can be discontinuous is $K^t_t$, which depends on $A'$.

At $\hat r_1$, the continuity of the metric Eq.\eqref{cont} requires

\begin{equation}
    \mathcal{K}_{\mbox{\tiny II}} = e^{A_{\mbox{\tiny I}}}(\hat r_1)\,(1-2\hat M\hat r_1^2)^{-\gamma}\ .
\end{equation}
The extrinsic curvature components are
\begin{equation}
    \hat K^t_t\big|_{\mbox{\tiny I}} = \left.\frac{\sqrt{h_1}}{2}\frac{A_{\mbox{\tiny I}}}{d\hat r}\right|_{\hat r_1} = \frac{2\hat M \hat r_1}{(3F_1-1)
    \sqrt{h_1}},
\end{equation}

\begin{equation}
     K^t_t\big|_{\mbox{\tiny II}} = -\gamma\frac{2\hat M \hat r_1}{\sqrt{h_1}},
\end{equation}

where $h_1 \equiv 1-2\hat M\hat r_1^2$ and $F_1\equiv F(\hat{r}_1)$. The jump in $\hat K^t_t$ is:

\begin{equation}
    [\hat K^t_t]\equiv \hat K^t_t\big|_{\mbox{\tiny II}} - \hat K^t_t\big|_{\mbox{\tiny I}}
    = -\frac{2\hat M \hat r_1}{\sqrt{h_1}}\left(\frac{\gamma(3F_1-1)+1}{3F_1-1}\right)\ ,
\end{equation}

and $[\hat K] = [\hat K^t_t]$ .
Hence, Eq.\eqref{smunu} demands that at the timelike surface $\hat r= \hat r_1$, matter with zero energy density and a tension given by 

\begin{equation}
    \hat\sigma_1 = -\hat S^\theta_\theta  
    =-\hat S^\phi_\phi=
    \frac{\hat M\hat r_1}{4\pi\sqrt{h_1}}\left(\frac{-\gamma(1-3F_1)+1}{1-3F_1}\right)\ ,
\end{equation} 

be present, which is consistent with the corresponding discontinuity in the radial pressure.

Metric continuity {at $\hat r_2$} yields
\begin{equation} 
    \mathcal{K}_{\mbox{\tiny II}} = e^{A_{III}(\hat r_2)}(1-2\hat M\hat r_2^2)^{-\gamma}\ .
\end{equation}
    
The relevant 
components of the extrinsic curvature are

\begin{equation}
    \hat K^t_t\big|_{\mbox{\tiny II}} = -\gamma\frac{2\hat M\hat r_2}{\sqrt{h_2}}\ , \qquad
    \hat K^t_t\big|_{\mbox{\tiny III}} = \frac{2\hat M\hat r_2}{(3F_2-1)\sqrt{h_2}}\ , 
\end{equation}

with $h_2\equiv 1-2\hat M\hat r_2^2$, $F_2\equiv F(r_2)$. Hence,

\begin{equation}
    [\hat K^t_t] = \hat K^t_t\big|_{\mbox{\tiny III}} - \hat K^t_t\big|_{\mbox{\tiny II}} = \frac{2\hat M\hat r_2}{\sqrt{h_2}}\left(\frac{\gamma(3F_2-1)+1}{3F_2-1}\right)\ .
\end{equation}
This gives zero energy density and tension
\begin{equation}
     \sigma_2 = \frac{\hat{M}\hat{r}_2}{4\pi\sqrt{h_2}}\left(\frac{\gamma(3F_2-1)+1}{3F_2-1}\right)\ ,
\end{equation}

which is negative for the case under consideration, and coincides in absolute value with the corresponding discontinuity in the pressure.

To build concrete realizations of these configurations, we use the following relation between the parameters $\hat r_1$ and $\hat r_2$ :

\begin{equation}
\label{equality}
    e^{A_{\mbox{\tiny I}}(\hat r_1)}(1-2\hat M\hat r_1^2)^{-\gamma} =e^{A_{\mbox{\tiny III}}(\hat r_2)}(1-2\hat M\hat r_2^2)^{-\gamma} ,
\end{equation}

which follows from the continuity of $g_{tt}$ at $\hat r_1$ and $\hat r_2$. To find the allowed values of $\hat r_2$, we build $e^{A_{\mbox{\tiny I}}(\hat r)}$ and $e^{A_{\mbox{\tiny III}}(r)}$, by numerically integrating $\frac{dA_{\mbox{\tiny I}}(\hat r)}{d\hat r}$ and $\frac{dA_{\mbox{\tiny III}}(\hat r)}{d\hat r}$, using the boundary conditions at $\hat r=0$ and $\hat r=1$, respectively. We then evaluate the lhs of Eq.\eqref{equality}
for a given value of $\hat r_1$, and find the values of $\hat r_2$ from Eq.\eqref{equality}. We consider values of $\hat r_1\lessapprox\hat r_d$.


\bibliography{apssamp}

\end{document}